\documentclass[journal,10pt]{IEEEtran}
\usepackage{bbm}
\usepackage{subfigure}
\usepackage{amsmath,amsfonts}
\usepackage{algorithmic}
\usepackage{algorithm}
\usepackage{array}
\usepackage{textcomp}
\usepackage{stfloats}
\usepackage{url}
\usepackage{verbatim}
\usepackage{graphicx}
\usepackage{color}
\usepackage{cite}
\usepackage{amsthm}
\usepackage{amssymb}
\usepackage{cleveref}
\usepackage{makecell}
\newcommand{\black}[1]{{\color{black}{#1}}}

\newcommand{\abs}[1]{\left\lvert #1\right\rvert}

\newcommand{\ceil}[1]{\left\lceil  #1\right\rceil}
\newcommand{\floor}[1]{\left\lfloor  #1\right\rfloor}
\newcommand{\Expect}[1]{\mathbb E\left[#1\right]}

\newcommand{\eff}{\mathrm{eff}}
\newcommand{\indic}[1]{\mathbbm{1}_{#1}}

\newcommand{\mya}{\mathrel{\overset{\makebox[0pt]{{\tiny(a)}}}{=}}}

\newcommand{\exc}{\mathrm{exc}}
\newcommand{\emb}{\text{emb}}
\newcommand{\fremb}{\widetilde{\free}}
\newcommand{\hybemb}{\widetilde{\hyb}}
\newcommand{\depemb}{\widetilde{\dep}}

\newcommand{\boldx}{\mathbf{x}}

\newcommand{\boldy}{\mathbf{y}}

\newcommand{\boldh}{\mathbf{h}}
\newcommand{\boldH}{\mathbf{H}}
\newcommand{\boldn}{\mathbf{n}}

\newcommand{\boldtau}{\boldsymbol{\tau}}
\newcommand{\boldnu}{\boldsymbol{\nu}}

\newcommand{\DD}{\mathrm{dd}}
\newcommand{\tx}{\mathrm{tx}}
\newcommand{\rx}{\mathrm{rx}}
\newcommand{\dep}{\mathrm{dep}}
\newcommand{\free}{\mathrm{free}}
\newcommand{\hyb}{\mathrm{hyb}}
\newcommand{\twistconvol}{*_{\sigma}}

\newcommand{\mask}{\mathrm{mask}}
\newcommand{\taup}{\tau_{\text{p}}}
\newcommand{\nup}{\nu_{\text{p}}}
\newcommand{\dd}{\text{dd}}
\theoremstyle{plain}

\theoremstyle{definition}

\theoremstyle{remark}

\graphicspath{{Figures/}{../Figures/}}
\begin{document}
\title{A Hybrid I/O Relation Estimation Scheme for Zak-OTFS Receivers}
\author{Sai Pradeep Muppaneni, Vineetha Yogesh, and A. Chockalingam \\
Department of ECE, Indian Institute of Science, Bangalore 560012 }
\maketitle

\begin{abstract}
In this paper, we consider the problem of estimating the delay-Doppler (DD) domain input-output (I/O) relation in Zak-OTFS modulation, which is needed for signal detection. Two approaches, namely, model-dependent and model-free approaches, can be employed for this purpose. The model-dependent approach requires explicit estimation of the physical channel parameters (path delays, Dopplers, and gains) to obtain the I/O relation. Such an explicit estimation is not required in the model-free approach, where the I/O relation can be estimated by reading off the samples in the fundamental DD period of the received pilot frame. Model-free approach has the advantage of acquiring fractional DD channels with simplicity. However, the read-off in the model-free approach provides an estimate of the effective channel only over a limited region in the DD plane but it does not provide an estimate for the region outside, and this can affect the estimation performance depending on the pulse shaping characteristics of the DD pulse shaping filter used. A poorly localized DD pulse shape leads to an increased degradation in performance. Motivated by this, in this paper, we propose a novel, yet simple, I/O relation estimation scheme that alleviates the above issue in the model-free approach. We achieve this by obtaining a coarse estimate of the effective channel outside the model-free estimation region using a novel model-dependent scheme and using this estimate along with the model-free estimate to obtain an improved estimate of the overall I/O relation. We devise the proposed estimation scheme for both exclusive and embedded pilot frames.
Our simulation results using Vehicular-A, TDL-A and TDL-C channel models with fractional DDs show that the proposed hybrid estimation approach achieves superior performance compared to the pure model-free approach. For example, at a bit error rate (BER) of $1.5 \times 10^{-3}$, the proposed scheme achieves an SNR gain of about 3.5 dB compared to the model-free approach for sinc filter in Veh-A channel, and this performance gain is achieved with a moderate increase in complexity (e.g., about $2.95\times 10^8$ and $8.1\times 10^8$ real operations for model-free approach and the proposed scheme, respectively). Also, the proposed scheme performs close to that of the sparse Bayesian learning (SBL) based estimation but at a significantly lesser complexity (e.g., about $8.3\times 10^{12}$ and $8.1\times 10^8$ real operations for the SBL based method and the proposed scheme, respectively).
\end{abstract}
\maketitle
\section{Introduction}
Ensuring reliable communication in high-mobility environments remains a significant challenge due to the resulting high Doppler spreads.  
Orthogonal time frequency space (OTFS) modulation overcomes this challenge through modulation in the delay-Doppler (DD) domain \cite{hadani}. Depending on how the conversion from DD domain to time domain is implemented, OTFS can be classified as multicarrier OTFS (MC-OTFS) or Zak-OTFS. Early research on OTFS has been based on the MC-OTFS approach \cite{hadani}-\cite{otfs_book}.
In this approach, the information-bearing DD domain signal is transformed into the time-frequency (TF) domain before being transformed into a time domain signal for transmission. Extensive work has been done on MC-OTFS \cite{best_reads}. An alternative approach that has recently gained attention is the Zak-OTFS approach \cite{bits},\cite{bits2}. Zak-OTFS uses the inverse Zak transform to directly convert the information-bearing DD domain signal into a time domain signal.
Although the Zak transform has been widely known in physics and signal processing research for quite some time \cite{Zak_existance},\cite{zak_transform},\cite{zak_sp}, it was first applied for modulation purposes in wireless communications by the authors in \cite{bits,bits2}.  
Zak-OTFS offers distinct advantages over MC-OTFS. It provides a formal mathematical framework using Zak theory to describe the Zak-OTFS waveform \cite{bits}. Its input-output (I/O) relation is a pure cascade of twisted convolution operations, unlike MC-OTFS where the I/O relation is a mix of linear convolution, multiplication, and twisted convolution operations (Table~3 in \cite{bits2}). This makes channel estimation in Zak-OTFS structurally simpler and predictable even in the presence of significant delay and Doppler spreads, and, as a consequence, the channel can be efficiently acquired and equalized \cite{bits2}. Further, Zak-OTFS has been shown to retain robustness under high Doppler spreads with estimated channels, compared to MC-OTFS (Fig.~18 in \cite{bits2}). Motivated by these advantages, research on Zak-OTFS is gaining momentum \cite{gopalam_2}-\cite{OTFSbook}.

\subsection*{Literature survey on Zak-OTFS}
Several works in the recent literature have contributed to the development and understanding of Zak-OTFS systems. In \cite{gopalam_2,gopalam_1}, the authors investigate the theoretically optimal receiver for Zak-OTFS and study its implementation through time-frequency windowing. Pulse shaping for DD communications is studied in \cite{zak_r2}. The work in \cite{tradeoff} introduces a framework to extend the region over which the condition for predictability holds. In \cite{fathima}, low-complexity search-based algorithms, such as likelihood ascent search and reactive tabu search when applied on top of linear minimum mean square error (LMMSE) equalization, are shown to bring the bit error rate (BER) performance close to the lower bound on maximum-likelihood detection performance. The work in \cite{z8} designs low-density parity check codes configured to Zak-OTFS by exploiting the reliability of channel estimates around the pilot location. Closed-form I/O relation expressions for  sinc and Gaussian filters under different receiver filter choices (identical, matched, and channel-matched) are derived in \cite{closed_form}. In \cite{new1}, a framework is developed to transform the orthogonal basis of Zak-OTFS pulsone waveforms into an orthonormal basis of spread carriers with low peak-to-average power ratio (PAPR). A compressive sensing-based approach for DD parameter estimation in integrated sensing and communication using Zak-OTFS modulation is proposed in \cite{new2}. The work in \cite{new3} investigates the effect of time-frequency windowing with different sidelobe behaviors and its impact on communication reliability. Implementation of Zak-OTFS as a low-complexity precoder over standard cyclic-prefixed orthogonal frequency division multiplexing (CP-OFDM) is reported in \cite{zak_over_ofdm}. A low-complexity frequency-domain equalization method is proposed in \cite{low_complex_equalization}, which reduces the detection complexity to quadratic order in the OTFS grid size. A differential scheme that alleviates the need for periodic pilot transmission is proposed in \cite{diff_eq}.  In \cite{per_carrier_eq}, a per-carrier precoding strategy is proposed, enabling separate equalization of each DD carrier with substantially lower complexity compared to joint equalization. In \cite{isac}, the authors design a superimposed spread pilot sequence that spans the entire OTFS grid along with the information symbols, thereby eliminating the need for any guard space and achieving $100\%$ spectral efficiency. In this paper, we focus on improving I/O relation estimation in Zak-OTFS which is crucial for achieving reliable data detection.

\subsection*{Problem and motivation}
The basic Zak-OTFS carrier waveform is a pulse in the DD domain which is a quasi-periodic localized function. Each DD pulse carries one data or pilot symbol, and there are multiple DD pulses (data/pilot symbols) in a frame. In order to limit the frame transmission within finite bandwidth and time duration, a DD domain pulse shaping filter is used at the transmitter. Sinc, Gaussian, and root raised cosine filters are commonly considered for this purpose \cite{bits2},\cite{tradeoff}-\cite{closed_form}. The DD characteristics of the pulse shaping filter (i.e., the main and sidelobe characteristics in the DD domain) influence the amount of DD domain aliasing due to the quasi-periodic replicas and the amount of inter-symbol interference (interference between data/pilot symbols). Consequently, the choice of this filter influences the receiver performance. 

In Zak-OTFS, the end-to-end DD domain I/O relation depends on the {\em effective channel}, consisting of the twisted convolution cascade of the transmit (Tx) DD filter, the physical channel, and the receive (Rx) DD filter. This paper addresses the problem of estimating this I/O relation, which is required for signal detection. Traditionally, this problem is solved using a model-dependent approach, where the parameters that characterize the physical channel (path delays, Dopplers, and gains) are explicitly estimated and used to obtain the I/O relation. This approach can be computationally intensive, particularly when channel delay and Doppler values are fractional. However, a unique advantage of Zak-OTFS is that, when operated in the crystalline regime where the delay period and Doppler period of the Zak transform are chosen to be larger than the maximum delay spread and maximum Doppler spread of the effective channel, respectively, the I/O relation can be obtained simply by reading off the received samples in the fundamental DD period. That is, no explicit estimation of the parameters of the physical channel is needed. Model-free approach, therefore, has the advantage of acquiring fractional DD channels with simplicity. However, the read-off in the model-free approach provides an estimate of the effective channel only over a limited region in the DD plane but it does not provide an estimate for the region outside, and this can affect the estimation performance depending on the pulse shaping characteristics of the DD filter used. A poorly localized pulse shape leads to significant performance degradation. For example, while the sinc filter has ideal main lobe characteristics (with nulls at the information DD grid points), it has high sidelobe levels which can affect 
the accuracy of the model-free I/O relation estimation\footnote{We will illustrate this in later sections through heat maps of the frames, and mean square error and bit error rate performance plots.}. This observation motivates the investigation of techniques that can alleviate this problem, which is the main focus of this paper. 

\subsection*{Solution approach and contributions}
Based on the above discussion, we see that an estimate of the effective channel outside the model-free estimation region is important. Model-dependent estimation can provide this, but at high complexity. We address this issue using a novel hybrid approach that combines the strengths of both model-free and model-dependent approaches. In the proposed solution, we retain the model-free read-off for its simplicity in acquiring fractional DDs and alleviate its performance shortfall using a low-complexity model-dependent add-on. The model-dependent add-on must be devised carefully so that it is computationally inexpensive yet gives a reasonable estimate of the channel parameters. One way to accomplish this is to obtain integer estimates of the delays and Dopplers using simple energy-based estimation and use them to obtain an estimate of the effective channel outside the model-free estimation region.
However, this is found to be inadequate to improve the overall I/O relation estimation performance. Therefore, we modify the energy-based model-dependent estimation to suit the needs of hybrid estimation in the context of fractional DDs. We first develop the proposed hybrid approach in the context of using an exclusive pilot frame, which consists of a pilot symbol at the center of the frame and zeros elsewhere. Exclusive pilot frames have the advantage of no data-to-pilot interference, but they incur throughput loss due to poor frame efficiency. Therefore, we next devise the proposed approach for the case of embedded pilot frames, where pilot and data symbols coexist and interfere with each other in the same frame. We evaluate the normalized mean square error (NMSE) performance of the proposed hybrid scheme and compare it with that of the pure model-free scheme. 
Our contributions in this paper can be summarized as follows.
\begin{itemize}
    \item We identify the performance bottleneck of model-free Zak-OTFS estimation under poorly localized DD filters such as the sinc filter, and analyze its limitations in capturing the full effective channel.
    \item We propose a hybrid estimation framework that augments the model-free read-off with a low-complexity model-dependent add-on, thereby achieving a more accurate I/O relation estimate.
    \item We design a simple yet effective model-dependent add-on based on energy-based path identification, tailored to fractional DD
    channels.
    \item We extend the scheme to embedded pilot structures, allowing estimation and detection in the presence of pilot-data interference.
    \item Our simulation results using Vehicular-A  \cite{vehA}, TDL-A and TDL-C \cite{3gpp} channel models with fractional DDs show that the proposed hybrid estimation approach achieves superior performance compared to the pure model-free approach. For example, at a bit error rate (BER) of $1.5 \times 10^{-3}$, the proposed scheme achieves an SNR gain of about 3.5 dB compared to the model-free approach for sinc filter in Vehicular-A channel, and this performance gain is achieved with a moderate increase in complexity (e.g., about $2.95\times 10^8$ and $8.1\times 10^8$ real operations for model-free approach and the proposed scheme, respectively). Also, the proposed scheme performs  close to that of the sparse Bayesian learning (SBL) based estimation \cite{off_grid} but at a significantly lesser complexity (e.g., about $8.3\times 10^{12}$ and $8.1\times 10^8$ real operations for the SBL based method and the proposed scheme, respectively).
\end{itemize}

The rest of the paper is structured as follows. Section \ref{sys_model} introduces the Zak-OTFS system model. Section \ref{IOR} presents the proposed hybrid estimation schemes for exclusive and embedded pilot frames. Section \ref{results} presents the NMSE and BER performance results and discussions. Conclusions are presented in Sec. \ref{conclusions}.

\section{System Model}
\label{sys_model}
\begin{figure}
\centering
\includegraphics[width=0.65\linewidth]{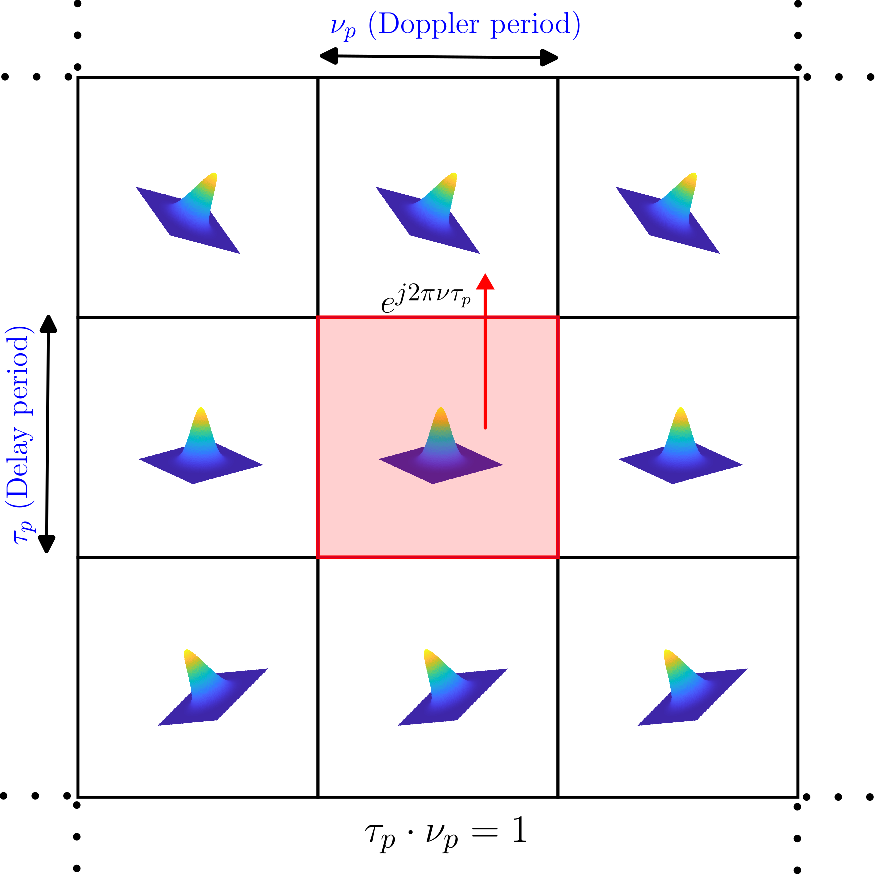}
\caption{Quasi-periodic DD domain pulse. Red box: fundamental DD period, ${\mathcal D}_0$. Quasi-periodic replicas in other boxes (a.k.a. DD domain aliases).}
\label{qp_dd_pulse}
\end{figure}
The basic information carrier in Zak-OTFS is a DD domain pulse which is a quasi-periodic function localized in a fundamental DD period, defined by a delay period $\tau_\text{p}$ and a Doppler period $\nu_\text{p}$ such that $\tau_\text{p}\nu_\text{p}=1$. The fundamental DD period is defined as 
$\mathcal{D}_0 = \{(\tau, \nu) \mid 0 \leq \tau < \tau_\mathrm{p}, 0 \leq \nu < \nu_\mathrm{p} \}$, where $\tau$ and $\nu$ denote the delay and Doppler variables, respectively. A two-dimensional signal $a(\tau,\nu)$ is quasi-periodic if it satisfies 
\begin{equation}
a(\tau+n\taup,\nu+m\nup) =e^{j2\pi \frac{n\nu}{\nup}}a(\tau,\nu), \ \ m,n\in \mathbb{Z},
\label{eqn:quasi-periodic}
\end{equation}
i.e., periodic in the Doppler domain and periodic up to a phase in the delay domain. A pictorial depiction of a quasi-periodic DD domain pulse is shown in Fig. \ref{qp_dd_pulse}, where the red box denotes the fundamental DD period ${\mathcal D}_0$ and the other boxes contain the quasi-periodic replicas (a.k.a. DD domain aliases). A quasi-periodic DD domain pulse, when viewed in the time domain, is a pulsone which is a time domain pulse train modulated by a frequency tone. The delay period $\tau_\text{p}$ is divided into $M$ delay bins and the Doppler period $\nu_\text{p}$ is divided into $N$ Doppler bins, and $MN$ information symbols are multiplexed on $MN$ DD pulses located in these $MN$ DD bins. $M$ and $N$ are chosen such that $MN = BT$, where $B$ and $T$ are the bandwidth and time duration, respectively, of a frame. That is, the resolution along the delay axis is $\frac{\tau_\text{p}}{M}= \frac{1}{B}$ and the resolution along the Doppler axis is $\frac{\nu_\text{p}}{N} = \frac{1}{T}$. In order to limit the bandwidth and time duration to $B$ and $T$, respectively, a DD domain pulse shaping filter is used at the transmitter.

\begin{figure}
\centering
\includegraphics[scale=0.45]{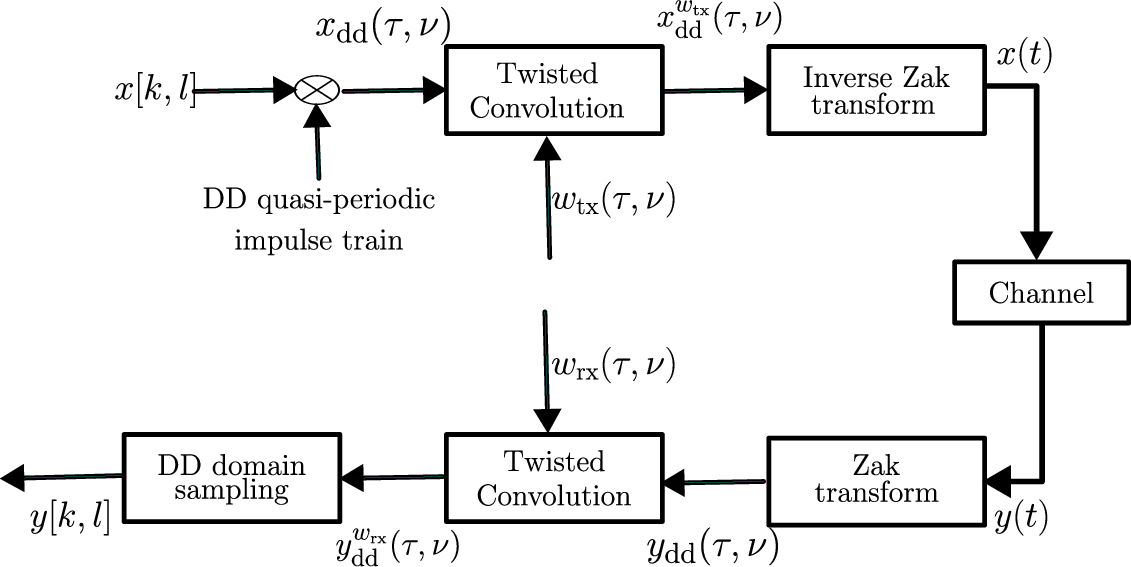}
\caption{Block diagram of the Zak-OTFS transceiver.}
\label{fig:block_diagram}
\end{figure}

Figure \ref{fig:block_diagram} shows the block diagram of a Zak-OTFS transceiver.
$MN$ information symbols, denoted by $x[k,l]$,  $k = 0,\ldots,M-1$, $l = 0,\ldots,N-1$, are drawn from a modulation alphabet $\mathbb{A}$, and
mounted on the information DD grid defined as $\Lambda_{\mathrm{dd}} \overset{\Delta}{=} \left\{\left(k\frac{\tau_\text{p}}{M},l\frac{\nu_\text{p}}{N}\right) \ \middle| \ k = 0,\ldots,M-1,\ l = 0,\ldots,N-1 \right\}$.
The $x[k,l]$s are encoded as per the following equation to obtain a quasi-periodic extension of the signal in the discrete DD domain:
\begin{equation}
x_{\mathrm {dd}}[k+nM,l+mN] = x[k,l]e^{j2\pi n \frac{l}{N}}, n,m\in\mathbb{Z}.
\label{quasi_per}
\end{equation}
The discrete quasi-periodic signal in (\ref{quasi_per}) is converted into a continuous quasi-periodic DD signal by mounting it on a continuous quasi-periodic DD impulse train, as
\begin{align}
x_{\mathrm{dd}}(\tau,\nu)   = &  \sum_{r,s\in\mathbb{Z}}x_{\mathrm {dd}}[r,s]\delta\left(\tau-r\tfrac{\taup}{M}\right)\delta\left(\nu-s\tfrac{\nup}{N}\right) \nonumber \\
\mya & \sum_{m,n\in\mathbb{Z}}\sum_{k=0}^{M-1}\sum_{l=0}^{N-1}x[k,l]e^{j2\pi \frac{nl}{N}} \nonumber \\
& \hspace{3mm}\delta\left(\tau-\tfrac{(k+nM)\taup}{M}\right)\delta\left(\nu-\tfrac{(l+mN)\nup}{N}\right),
\label{eqn:x_DD}
\end{align}
where $\delta(\cdot)$ denotes the Dirac delta function, and step (a) is obtained by substituting the variables \(r\) and \(s\) by $k+nM$ and $l+mN$, respectively. The signal in \eqref{eqn:x_DD} is then time and bandwidth limited to $T=N\taup$ and $B=M\nup$, respectively, by filtering through the DD domain Tx filter $w_{\mathrm{tx}}(\tau,\nu)$,
as \cite{bits2}
\begin{equation}
\label{Tx_filtering}
x_{\mathrm{dd}}^{w_{\mathrm{tx}}}(\tau,\nu) = w_{\mathrm{tx}}(\tau,\nu) *_{\sigma} x_{\mathrm{dd}}(\tau,\nu),
\end{equation} 
where $*_{\sigma}$ denotes the twisted convolution operation. Twisted convolution between two functions $a(\tau,\nu)$ and $b(\tau,\nu)$ is defined as 
\begin{eqnarray}
a(\tau,\nu)*_{\sigma}b(\tau,\nu) & \hspace{-2mm} = & \hspace{-2mm} \int_{-\infty}^{\infty}\int_{-\infty}^{\infty} a(\tau',\nu')b(\tau-\tau',\nu-\nu') \nonumber \\
& \hspace{-2mm} & \hspace{-2mm} e^{j2\pi\nu'(\tau-\tau')}d\tau'd\nu'.
\label{eqn:tc}
\end{eqnarray}
Also, twisted convolution operation between the quasi-periodized DD signal $x_\text{dd}(\tau,\nu)$ and the non-quasi-periodic DD filter/function $w_\text{tx}(\tau,\nu)$ in (\ref{Tx_filtering}) preserves the quasi-periodicity at the output (Appendix 2.G, \cite{OTFSbook}), i.e., the DD signal $x_\text{dd}^{w_\text{tx}}(\tau,\nu)$ is quasi-periodic. This quasi-periodicity of $x_\text{dd}^{w_\text{tx}}(\tau,\nu)$ ensures the existence of the corresponding time domain (TD) signal through inverse Zak transform operation. 
Accordingly, the TD transmit signal is obtained using inverse Zak transform\footnote{The inverse Zak transform of a DD domain signal $a(\tau,\nu)$ is defined as $\mathcal{Z}_{t}^{-1} \left(a(\tau,\nu)\right) \overset{\Delta}{=} \sqrt{\tau_\mathrm{p}} \int_0^{\nu_\mathrm{p}} a(t,\nu) d\nu.$} as 
\begin{equation}
\label{s_td}
x(t) = \mathcal{Z}^{-1}_{t}\left(x_{\mathrm{dd}}^{w_{\mathrm{tx}}}(\tau,\nu)\right) = \sqrt{\taup}\int_0^{\nup}x_{\mathrm{dd}}^{w_{\mathrm{tx}}}(t,\nu)d\nu,
\end{equation}
where $\mathcal{Z}_t^{-1}(.)$ denotes the inverse Zak transform operation.
The transmitted signal $x(t)$ passes through a doubly-selective channel, whose DD domain impulse response is given by
\begin{equation}
\label{phy_ch}
h(\tau,\nu) = \sum_{i=1}^P h_i \delta(\tau-\tau_i)\delta(\nu-\nu_i),
\end{equation}
where $h_i,\tau_i,$ and $\nu_i$ are the channel gain, delay, and Doppler of the $i$th path, respectively, and $P$ is the number of paths. The received TD signal at the receiver is given by
\begin{eqnarray}
\label{r_td}
y(t) & \hspace{-2mm} = & \hspace{-2mm} \int\int h(\tau,\nu) x(t-\tau)e^{j2\pi\nu(t-\tau)}d\tau d\nu + n(t) \nonumber \\
& \hspace{-2mm} = & \hspace{-2mm}
\sum_{i=1}^Ph_ix(t-\tau_i)e^{j2\pi \nu_i(t-\tau_i)}+n(t),
\end{eqnarray}
where $n(t)$ is the additive white Gaussian noise with one-sided power spectral density $N_0$ W/Hz.
The received TD signal is converted to a DD domain signal using Zak transform\footnote{The Zak transform of a time domain signal $a(t)$ is given by $\mathcal{Z}_{t} \left(a(t)\right) \overset{\Delta}{=} \sqrt{\tau_\mathrm{p}}  \sum_{k \in \mathbb{Z}} a(\tau+k\tau_\mathrm{p}) e^{-j2\pi\nu k\tau_\mathrm{p}}$.} as
\begin{equation}
y_{\mathrm{dd}}(\tau,\nu) = \mathcal{Z}(y(t))
= \sqrt{\taup}\sum_{k\in \mathbb{Z}}y(\tau+k\taup)e^{-j2\pi k\nu\taup}.
\label{ydd}
\end{equation}
The DD signal $y_{\mathrm{dd}}(\tau,\nu)$ in (\ref{ydd}) can be written as 
\begin{eqnarray}
\hspace{-6mm}
y_{\mathrm{dd}}(\tau,\nu) & \hspace{-2mm} = & \hspace{-2mm} h(\tau,\nu)\twistconvol x_{\mathrm{dd}}^{w_{\tx}}(\tau,\nu)+n_{\mathrm{dd}}(\tau,\nu) \nonumber \\
& \hspace{-17mm} = & \hspace{-10mm} h(\tau,\nu)\twistconvol\left(w_{\tx}(\tau,\nu)\twistconvol x_{\mathrm{dd}}
(\tau,\nu)\right)+n_{\mathrm{dd}}(\tau,\nu) \nonumber \\
& \hspace{-17mm} \mya & \hspace{-10mm} \left(h(\tau,\nu)\twistconvol w_{\tx}(\tau,\nu)\right)\twistconvol x_{\mathrm{dd}}(\tau,\nu)+n_{\mathrm{dd}}(\tau,\nu),
\end{eqnarray}
where step (a) is due to the associative property of twisted convolution \cite{OTFSbook}, and $n_{\mathrm{dd}}(\tau,\nu)$ is the Zak transform of $n(t)$. The DD signal $y_{\mathrm{dd}}(\tau,\nu)$ is filtered using a receive DD domain filter $w_{\mathrm{rx}}(\tau,\nu)$ matched to the transmit DD filter, i.e., 
\begin{equation}
\label{rx_fil}
w_{\mathrm{rx}}(\tau,\nu) = w_{\mathrm{tx}}^{*}(-\tau,-\nu)e^{j2\pi \tau\nu}.
\end{equation}
The output of the receive DD filter is given by
\begin{eqnarray}
y_{\mathrm{dd}}^{w_{\rx}}(\tau,\nu)& \hspace{-2mm} = & \hspace{-2mm} w_{\rx}(\tau,\nu)\!\twistconvol\! y_{\mathrm{dd}}(\tau,\nu) \nonumber \\
& \hspace{-2mm} = & \hspace{-2mm} w_{\rx}(\tau,\nu)\!\twistconvol\! \left(h(\tau,\nu)\!\twistconvol\! w_{\tx}(\tau,\nu)\right)\!\twistconvol\! x_{\mathrm{dd}}(\tau,\nu) \nonumber \\
& \hspace{-2mm} & \hspace{-2mm}+w_{\rx}(\tau,\nu)\!\twistconvol\! n_{\mathrm{dd}}(\tau,\nu) \nonumber \\
& \hspace{-2mm} = & \hspace{-2mm} \underbrace{\left(w_{\rx}(\tau,\nu)\twistconvol h(\tau,\nu)\!\twistconvol\! w_{\tx}(\tau,\nu)\right)}_{\overset{\Delta}{=}\ h_{\eff}(\tau,\nu)}\!\twistconvol x_{\mathrm{dd}}(\tau,\nu) \nonumber \\
& \hspace{-2mm} & \hspace{-2mm} 
+\underbrace{w_{\rx}(\tau,\nu)\twistconvol n_{\mathrm{dd}}(\tau,\nu)}_{\overset{\Delta}{=}\ n_{\mathrm{dd}}^{w_{\rx}}(\tau,\nu)},
\label{eqn:y_DD}
\end{eqnarray}
where $h_{\mathrm{eff}}(\tau,\nu)$ is the effective channel given by the twisted convolution cascade of the Tx DD filter, physical DD channel, and Rx DD filter, given by
\begin{equation}
h_{\mathrm{eff}}(\tau,\nu) =
w_{\rx}(\tau,\nu)\twistconvol h(\tau,\nu)\twistconvol w_{\tx}(\tau,\nu),
\label{hdd}
\end{equation}
and $n_{\mathrm{dd}}^{w_{\mathrm{rx}}}(\tau,\nu)$ is the filtered noise in the DD domain. The filtered DD signal $y_{\mathrm{dd}}^{w_{\rx}}(\tau,\nu)$ is sampled at $\tau=\frac{k'\taup}{M},\nu=\frac{l'\nup}{N}$, $k'=0,1,\ldots,M-1$, $l'=0,1,\ldots,N-1$,
i.e.,
\begin{equation}
y[k',l']=y_{\mathrm{dd}}^{w_{\rx}}(\tau,\nu)\Big\vert_{\tau=\frac{k'\taup}{M},\nu=\frac{l'\nup}{N}}.
\label{eqn:y_DD_discrete}
\end{equation}
Note that it suffices to sample $y_{\mathrm{dd}}^{w_{\rx}}(\tau,\nu)$ in the fundamental period $\mathcal{D}_0$ as the signal $y_{\mathrm{dd}}^{w_{\rx}}(\tau,\nu)$ is quasi-periodic in nature. Substituting \eqref{eqn:x_DD} in \eqref{eqn:y_DD} and solving further, \eqref{eqn:y_DD_discrete} can be written as (see Appendix A for proof)
\begin{eqnarray}
\label{eqn:sys_model_non_vec}
\hspace{-2mm}
y[k',l']
& \hspace{-2mm} = & \hspace{-3mm}
\sum_{m,n\in\mathbb{Z}} \sum_{k=0}^{M-1}\!\sum_{l=0}^{\ N-1}\!
h_{\eff}[k'-k-nM,l'-l-mN] \nonumber \\
& \hspace{-2mm}  & \hspace{-3mm} x[k,l]e^{j2\pi\frac{nl}{N}}e^{j2\pi\frac{(l'-l-mN)(k+nM)}{MN}} + n[k',l'],
\end{eqnarray}
where $n[k',l']=n_{\dd}^{w_{\rx}}(k'\frac{\taup}{M},l'\frac{\nup}{N})$ and the discrete samples of the effective channel are obtained as $h_{\eff}[r,s]=h_{\eff}(\tau=r\frac{\taup}{M},\nu=s\frac{\nup}{N})$.
Writing the $y[k',l']$ samples as a vector, the end-to-end DD domain I/O relation can be expressed in matrix-vector form as
\begin{equation}
\boldy=\boldH\boldx+\boldn,
\label{eqn:sys_mod_vec}
\end{equation}
where 
$\mathbf{x},
\mathbf{y},
\mathbf{n} 
\in\mathbb{C}^{MN\times 1}$, such that  $\mathbf{x}[lM+k]=x[k,l]$, $\mathbf{y}[lM+k]=y[k,l]$, $\mathbf{n}[lM+k]=n[k,l]$, and
$\mathbf{H}\in\mathbb{C}^{MN\times MN}$ is the DD channel matrix such that
\begin{align}  \boldH[l'M+k',lM+k]&=\hspace{-3mm}\sum_{m,n\in\mathbb{Z}}\hspace{-2mm}h_{\eff}[k'\!-\!k\!-\!nM,l'\!-\!l\!-\!mN]
\nonumber\\
&\hspace{5mm}e^{j2\pi\frac{nl}{N}}
e^{j2\pi\frac{(l'-l-mN)(k+nM)}{MN}},
\label{eqn:H_vec}
\end{align}
where $k',k=0,1,\ldots,M-1$ and $l',l=0,1,\ldots,N-1$.

\textit{Remark:} In computing the entries of ${\bf H}$ 
as per (\ref{eqn:H_vec}), we limit the summation to $\abs{m}\leq m_{\max}$, $\abs{n}\leq n_{\max}$, and $m_{\max}=n_{\max}=2$ is found to give adequate accuracy.

A Zak-OTFS system is said to be operating in the {\em crystalline region} if the {\em crystallization condition} is met. The crystallization condition is said to be met if the delay spread of the effective channel is less than the delay period $\tau_{\mathrm{p}}$ and the Doppler spread of the effective channel is less than the Doppler period $\nu_{\mathrm{p}}$. By operating the system in the crystalline region, the leakage of the quasi-periodic replicas into the fundamental region is limited. Because of this, the I/O relation, which is required for equalization/detection, is predictable and non-fading \cite{bits}, \cite{bits2}. This allows the estimation of the I/O relation through a simple read-off in the fundamental region. 

We consider sinc and Gaussian pulse shaping filters. For sinc filter, $w_{\tx}(\tau,\nu)$ is given by \begin{equation} w_{\tx}(\tau,\nu)=\sqrt{BT}\mathrm{sinc}(B\tau)\mathrm{sinc}(T\nu). \end{equation} The expressions for $h_{\eff}[k,l]$ and noise covariance for the sinc filter are given in \eqref{eqn:sinc_heff} and \eqref{eqn:sinc_cov}, respectively \cite{closed_form}, where $\indic{\{.\}}$ denotes the indicator function, \(\ceil{.}\) and $\floor{.} $ denote the ceil and floor operators, respectively. For Gaussian filter, $w_{\tx}(\tau,\nu)$ is given by \begin{equation} w_{\tx}(\tau,\nu)=\left(\frac{2\alpha_{\tau}B^2}{\pi}\right)^{\frac{1}{4}}e^{-\alpha_\tau B^2\tau^2} \left(\frac{2\alpha_\nu T^2}{\pi}\right)^{\frac{1}{4}}e^{-\alpha_\nu T^2\nu^2}. \end{equation} The parameters $\alpha_{\tau}$ and $\alpha_\nu$ are chosen to be 1.584 which ensures that $99\%$ of the frame energy is contained in $B$ in frequency domain and $T$ in time domain. The expressions for $h_{\eff}[k,l]$ and noise covariance for the Gaussian filter are given in \eqref{eqn:gauss_heff} and \eqref{eqn:gauss_cov}, respectively \cite{closed_form}. In computing \eqref{eqn:gauss_cov}, taking $q_1$ and $q_2$ values from $-20$ to $20$ is found to be sufficient for accurate computation of \eqref{eqn:gauss_cov}. 

\begin{figure}
\centering
\includegraphics[width=0.95\linewidth]{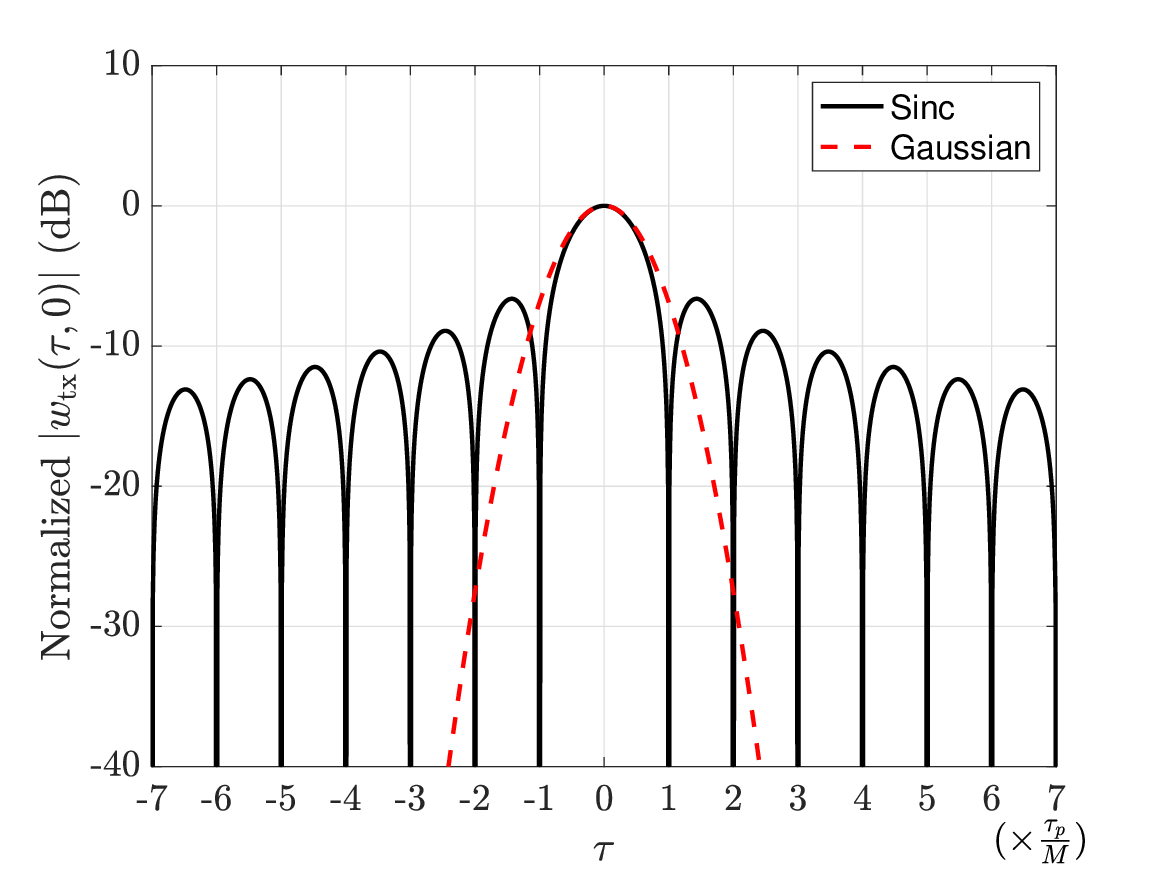}
\caption{Sinc and Gaussian pulse shapes.}
\label{fig:filters}
\end{figure}

\begin{figure*}[t]
For sinc filter:
   {\small \begin{equation}
    h_{\eff}[k,l]\hspace{-1mm}=\hspace{-1mm}\sum_{i=1}^P h_ie^{j\pi \left(\tfrac{kl}{MN}-\tau_i\nu_i\right)}\left(1-\abs{\tfrac{k\taup}{MT}}\right)\left(1-\abs{\tfrac{\nu_i}{B}}\right)\text{sinc}((\tfrac{l\nup}{N}-\nu_i)(T-|\tfrac{k\taup}{M}|))\indic{\{\abs{\nu_i}<B\}} \mathrm{sinc}((\tfrac{k\taup}{M}-\tau_i)(B-|\nu_i|))\indic{\{|k\frac{\taup}{M}|<T\}}.
    \label{eqn:sinc_heff}
\end{equation}
\begin{equation}
\Expect{n_{\DD}^{w_{\rx}}[k_1,l_1]{n_{\DD}^{w_{\rx}}}^H[k_2,l_2]}= N_0\tfrac{\taup}{T}\hspace{-3mm}\sum_{q_1=\ceil{-\frac{N}{2}-\frac{k_1}{M}}}^{\floor{\frac{N}{2}-\frac{k_1}{M}}}
\sum_{{q_2=\ceil{-\frac{N}{2}-\frac{k_2}{M}}}}^{{\floor{\frac{N}{2}-\frac{k_2}{M}}}} \hspace{-4mm} \text{sinc}\big(\tfrac{B\taup}{M}(k_1-k_2)+(q_1-q_2)\taup)\big) 
e^{-j2\pi(q_1\frac{l_1}{N}-q_2\frac{l_2}{N})}.
\label{eqn:sinc_cov}
\end{equation}}
\hrule
\vspace{1mm}
For Gaussian filter:
{\small
\begin{equation}
    h_{\eff}[k,l]=\sum_{i=1}^{P}h_ie^{-\frac{1}{2}\left(\alpha_\tau B^2(\tau_i-\frac{k\taup}{M})^2+\alpha_\nu T^2(\nu_i-\frac{l\nup}{N})^2\right)}e^{-\frac{\pi^2}{2}\left(\frac{\nu_i^2}{\alpha_\tau B^2}+\frac{k^2\taup^2}{M^2\alpha_{\nu}T^2}\right)}e^{-j\pi\left(\tau_i\nu_i-\frac{kl}{MN}\right)}.
    \label{eqn:gauss_heff}
\end{equation} 
\begin{equation}
    \Expect{n_{\DD}^{w_{\rx}}[k_1,l_1]{n_{\DD}^{w_{\rx}}}^H[k_2,l_2]}=N_0\tfrac{\taup}{T}\sqrt{\tfrac{2\pi}{\alpha_{\nu}}}\hspace{-1mm}\sum_{q_1,q_2\in\mathbb{Z}}\hspace{-2mm}e^{-j2\pi\left(\frac{q_1l_1-q_2l_2}{N}\right)}e^{-\pi^2\frac{\taup^2}{\alpha_{\nu}T^2}\left((q_1+\frac{k_1}{M})^2+(q_2+\frac{k_2}{M})^2\right)}e^{-\frac{\alpha_{\tau}B^2}{2}\left((k_1-k_2)\frac{\taup}{M}+(q_1-q_2)\taup\right)^2}.
    \label{eqn:gauss_cov}
\end{equation}}
\hrule

\end{figure*}

Apart from being widely considered filters in the Zak-OTFS literature, the choice of sinc and Gaussian filters is motivated by their contrasting DD domain characteristics and corresponding effect on I/O relation estimation and detection performance.  Figure~\ref{fig:filters} shows the sinc and Gaussian pulse shapes. The sinc filter has nulls at integer multiples of $\left(\tfrac{1}{B},\tfrac{1}{T}\right)=\left(\tfrac{\tau_{\mathrm p}}{M},\tfrac{\nu_{\mathrm p}}{N}\right)$, i.e., at the DD grid locations where information symbols are placed. This property offers the benefit of no inter-symbol interference at the information grid points at which receiver sampling is done. This is attractive from an equalization/detection viewpoint. However, a drawback with sinc filter is its high sidelobes. About $18.5\%$ of the energy in sinc filter lies in the sidelobes (energy outside the region $\big[-\tfrac{\tau_{\mathrm p}}{M},\tfrac{\tau_{\mathrm p}}{M}\big)\times\big[-\tfrac{\nu_{\mathrm p}}{N},\tfrac{\nu_{\mathrm p}}{N}\big)$). This high sidelobe levels leave the sinc pulse less localized in the DD domain, which is detrimental from an I/O relation estimation viewpoint. In contrast, the Gaussian pulse is much better localized in the DD domain with only about $2.35\%$ of the energy lying outside the region $\big[-\tfrac{\tau_{\mathrm p}}{M},\tfrac{\tau_{\mathrm p}}{M}\big)\times\big[-\tfrac{\nu_{\mathrm p}}{N},\tfrac{\nu_{\mathrm p}}{N}\big)$, and this compactness renders the Gaussian filter attractive for I/O relation estimation. However, Gaussian pulse does not have nulls at the information grid points, which is detrimental for equalization/detection. That is, sinc filter offers good detection but poor estimation, while Gaussian filter offers good estimation but poor detection. This trade-off is quantitatively demonstrated in Fig.~\ref{fig:new} in the next section.

\begin{figure}
    \centering
    \subfigure[Exclusive pilot frame.]{\label{fig:exc_pil}\includegraphics[width=0.39\linewidth]{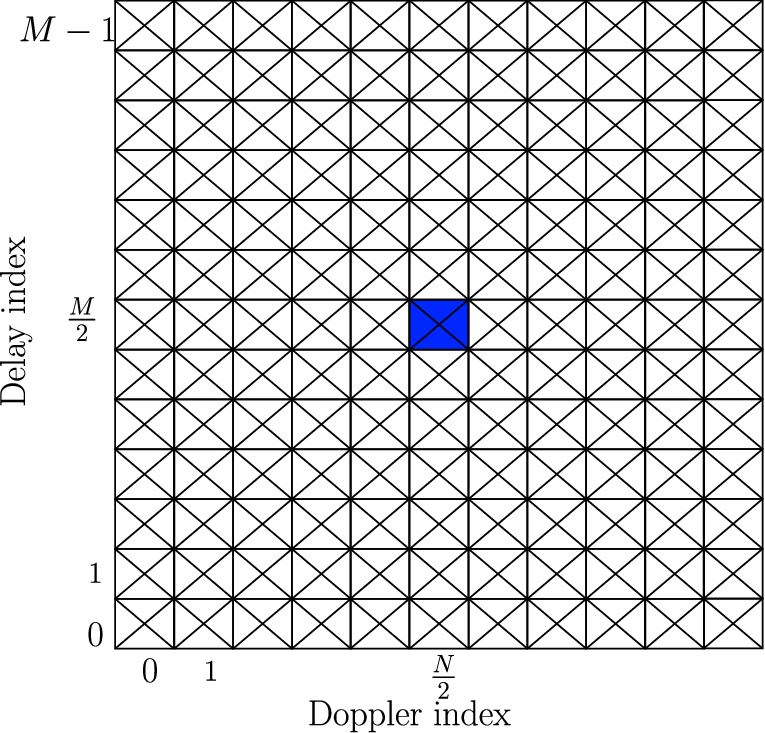}}
    \hfill
    \subfigure[Embedded frame.]{\label{fig:emb_pil}\includegraphics[width=0.5839\linewidth]{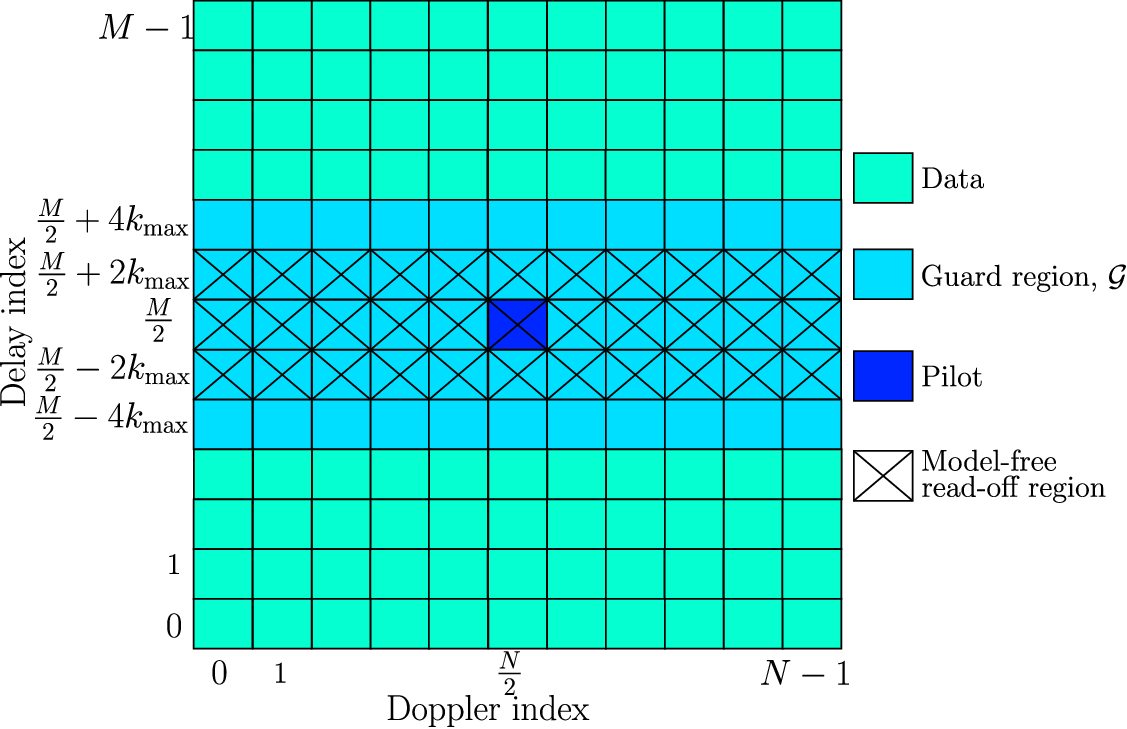}}
     \caption{Pilot frames used for estimating the channel in Zak-OTFS.}
    \label{fig:embedded}
\end{figure}

\section{Proposed hybrid I/O relation estimation}
\label{IOR}
In order to detect the data symbols (i.e., {\bf x} vector), the receiver needs to know the  I/O relation matrix $\boldH$. To obtain an estimate of the ${\bf H}$ matrix, we consider two types of frames, namely, exclusive pilot frame and embedded pilot frame. Exclusive pilot frame, which has only one pilot symbol and zeros elsewhere in the frame (Fig. \ref{fig:exc_pil}), has the advantage of simplicity at the cost of frame efficiency. Embedded pilot frame, which consists of a pilot symbol and data symbols with guard region in between (Fig. \ref{fig:emb_pil}), is more throughput efficient but has to deal with pilot-data interference. In the following subsections, we present the proposed I/O relation estimation scheme for exclusive and embedded pilot frames.

\subsection{Estimation for exclusive pilot frame}
\label{sec:hybrid}
In an exclusive pilot frame, a pilot symbol $\alpha\in\mathcal{C}$ is placed at the center of the frame and zeros elsewhere (see Fig. \ref{fig:exc_pil}), i.e., for $k=0,1,\cdots,M-1, l=0,1,\cdots,N-1$, 
\begin{equation}
    x_{\exc}[k,l]=\begin{cases}
        \alpha\hspace{10mm} \text{if }(k,l)=(\frac{M}{2},\frac{N}{2})\\
        0 \hspace{10mm} \text{otherwise.}
    \end{cases}
    \label{eqn:exclusive_pilot}
\end{equation}
Two approaches for I/O relation estimation, namely, model-dependent approach and model-free approach, are possible, which are introduced below. The estimated I/O relation using the exclusive pilot frame is used for the detection of data frames sent during the same spatial coherence interval.

\subsubsection{Model-dependent estimation}
In the model-dependent approach, the parameters of the physical channel $h_\mathrm{}(\tau,\nu)$, i.e., $\{\tau_i$, $\nu_i$, $h_i$\}, $i=1,\ldots,P$, are explicitly estimated and $h_\mathrm{eff}[k,l]$ is computed using \eqref{hdd}, and subsequently the matrix $\mathbf{H}_\mathrm{}$  is obtained using \eqref{eqn:H_vec}. This approach is computationally complex, particularly for channels with fractional DDs, which are more practical.

\subsubsection{Model-free estimation}
In the model-free approach, an estimate of the ${\bf H}$ matrix can be obtained without explicit estimation of the parameters of the physical channel model as follows. Using \eqref{eqn:exclusive_pilot} in \eqref{eqn:sys_model_non_vec}, the received exclusive pilot frame signal can be written as 
\begin{eqnarray}
\label{eqn:y_pil}
y_{\exc}[k',l'] & \hspace{-2mm} = & \hspace{-2mm}\sum_{m,n\in\mathbb{Z}}\hspace{-2mm}\alpha h_{\eff}[k'-\frac{M}{2}-nM,l'-\frac{N}{2}-mN] \nonumber\\
&\hspace{-2mm} & \hspace{-2mm} e^{j2\pi\frac{nl}{N}}e^{j2\pi\frac{(l'-\frac{N}{2}-mN)(k'+mN)}{MN}} + n[k',l'].
\label{eqn:excl_pil_rsp}
\end{eqnarray}
In \eqref{eqn:excl_pil_rsp}, the summation over $m,n\in \mathbb{Z}\backslash\{0\}$ denotes the leakage of pilot signal from the non-fundamental periods into the fundamental period. 
Since $y_{\exc}[k',l']$ is quasi-periodic, only the samples falling within the fundamental period constitute the unique observable samples corresponding to the transmitted pilot frame (other samples in the non-fundamental periods are replicas). In the crystalline regime of operation, where the delay and Doppler spreads of the effective channel are less than the delay and Doppler periods, respectively, the leakage from the neighboring grids can be small. 
Consequently, by restricting $m$ and $n$ to $0$, i.e., the fundamental period,   
\eqref{eqn:y_pil} in the noiseless case
can be written as  
\begin{equation}
    y_{\exc}[k',l'] 
     = \alpha h_{\eff}\Big[k'-\frac{M}{2},l'-\frac{N}{2}\Big]e^{j\pi\frac{l'-\frac{N}{2}}{N}}.
    \label{eqn:y_exc}
\end{equation}
By substituting $k'-\frac{M}{2}=a$ and $l'-\frac{N}{2}=b$, the above equation becomes
\begin{equation}
    h_{\eff}[a,b]=y_{\exc}\Big[a+\frac{M}{2},b+\frac{N}{2}\Big]\frac{e^{-j\pi\frac{b}{N}}}{\alpha},
    \label{eqn:m_free}
\end{equation}
which is taken as the model-free estimate $\hat{h}_{\eff}^{\free}[a,b]$ of the effective channel tap $h_{\eff}[a,b]$. Since the range of observations of $y_{\exc}[k',l']$ is from $0\leq k'<M, 0\leq l'<N$, the range of $h_{\eff}[a,b]$ obtained from \eqref{eqn:m_free} is from $-\frac{M}{2}\leq a<\frac{M}{2},-\frac{N}{2}\leq b<\frac{N}{2}$. For all values other than this, the estimate is set to be zero, i.e., the model-free estimates $\hat{h}_{\eff}^{\free}[a,b]$ are obtained as
\begin{equation}
    \hat{h}_{\eff}^{\free}[a,b]\hspace{-1mm}=\hspace{-1mm}\begin{cases}
        y_{\exc}[a+\frac{M}{2},b+\frac{N}{2}]\frac{e^{-j\pi\frac{b}{N}}}{\alpha}, \hspace{2mm} \text{if} -\frac{M}{2}\leq a <\frac{M}{2},\\
        \hspace{46mm}-\frac{N}{2}\leq b<\frac{N}{2},\\
        0, \hspace{39mm} \text{otherwise.}
    \end{cases}
    \label{eqn:m_free_2}
\end{equation}
Note that the $y_{\exc}[k',l']$ read-off region, denoted by $\mathcal{R}_{\exc}$,
is the entire pilot frame, i.e.,
$\mathcal{R}_{\exc}=\{0,1,\cdots,M-1\}\times\{0,1,\cdots,N-1\}$. 
The corresponding model-free estimation region for $\hat{h}_{\eff}^{\free}[a,b]$, denoted by $\mathcal{F}_{\exc}$, is $\mathcal{F}_{\exc}=\{-\frac{M}{2},\cdots,\frac{M}{2}-1\}\times\{-\frac{N}{2},\cdots,\frac{N}{2}-1\}$.
Using the estimated channel taps $\hat{h}_{\eff}^{\free}[a,b]$
from \eqref{eqn:m_free_2} 
in \eqref{eqn:H_vec}, we obtain the model-free estimate of the ${\bf H}$ matrix for signal detection.
\begin{figure}
    \centering
    \subfigure[Gaussian filter.]{\includegraphics[width=0.49\linewidth]{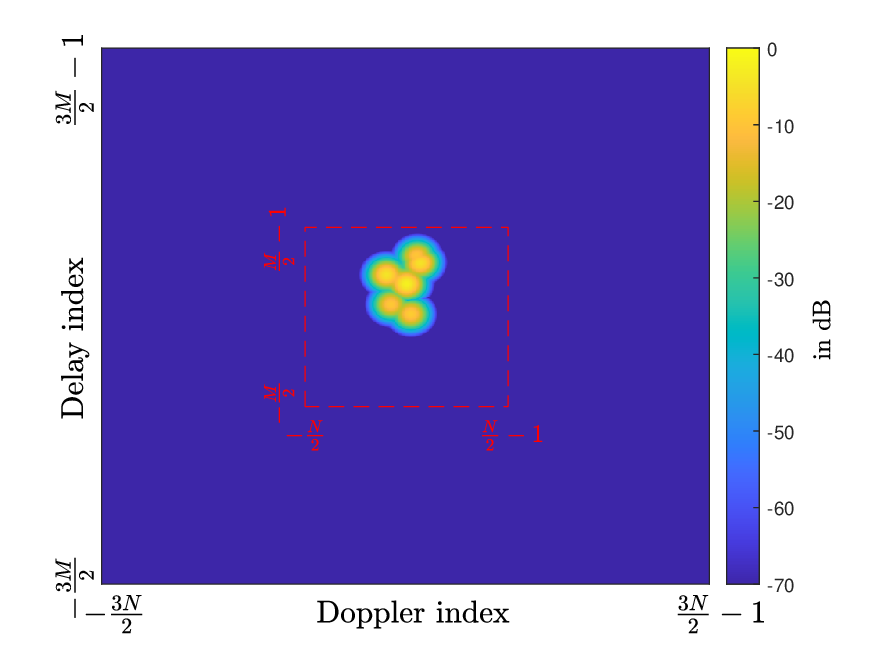}\label{fig:gauss_h_eff}}
    \subfigure[Sinc filter.]{\label{fig:sinc_h_eff}\includegraphics[width=0.49\linewidth]{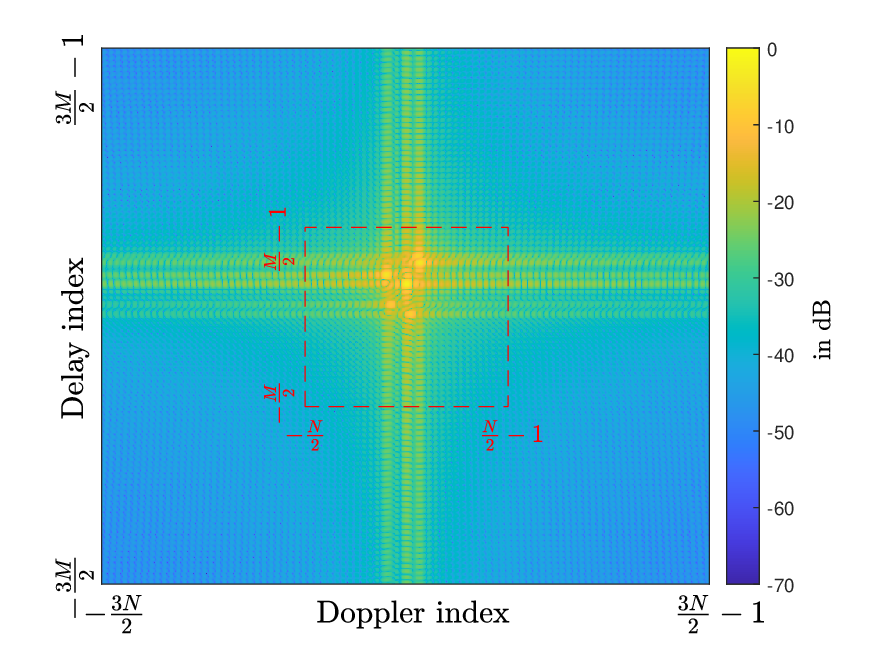}}
     \caption{
     Heatmap of $\abs{h_{\eff}}$
     for Gaussian and sinc filters with exclusive pilot frame. 
     The highlighted box in red corresponds to the model-free estimation region $\mathcal{F}_{\exc}$. $\abs{h_{\eff}}$
     is well contained within this region with Gaussian filter while it is not with sinc filter.}
    \label{fig:h_filter_eff}
\end{figure}

\subsubsection{Rationale for the proposed hybrid estimation}
\begin{figure}
    \centering
    \subfigure[Quasi-periodic received exclusive pilot signal in \eqref{eqn:y_exc}. Highlighted red box shows the fundamental period.]{\label{fig:rx_exc_pil}\includegraphics[width=0.49\linewidth]{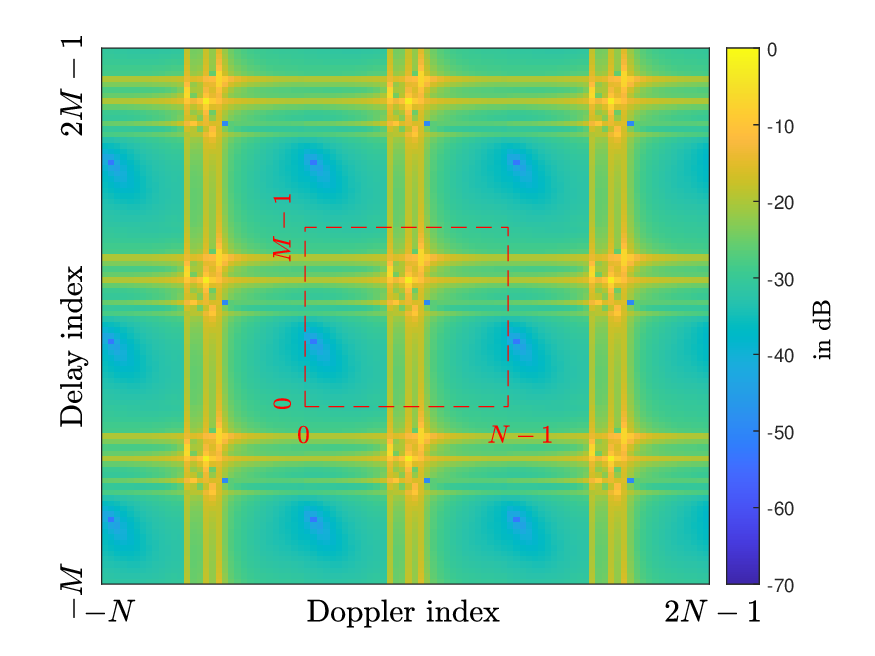}}
    \subfigure[Heatmap of model-free estimates $|{\hat{h}_{\eff}^{\free}}|$ as per \eqref{eqn:m_free_2}.] 
{\label{fig:mod_free_exc_pil}\includegraphics[width=0.49\linewidth]{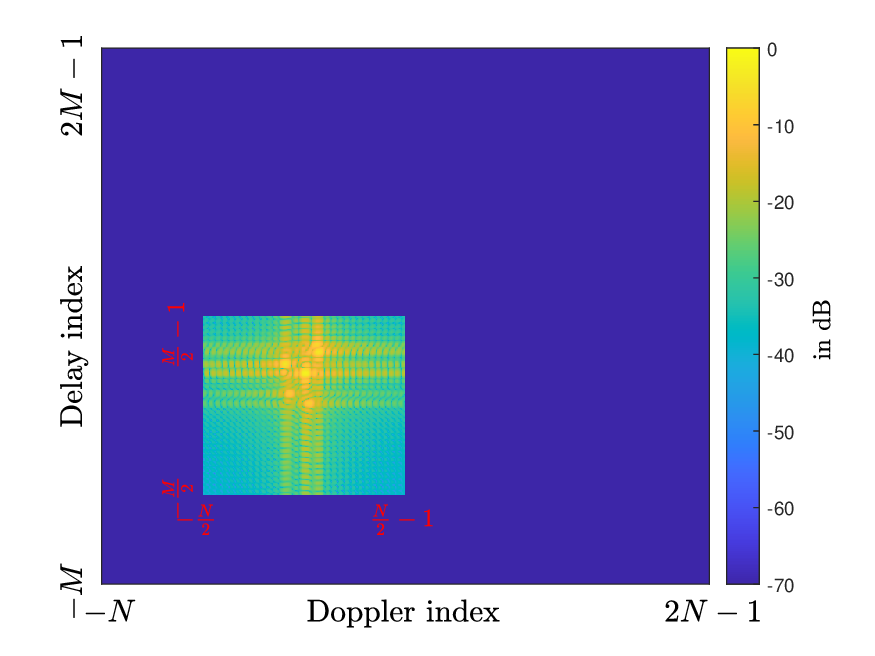}}\\
     \subfigure[Heatmap of true $\abs{h_{\eff}}$ as per \eqref{hdd}. Highlighted red box shows the model-free estimation region $\mathcal{F}_{\exc}$.]{\label{fig:true_h_eff_sinc}\includegraphics[width=0.49\linewidth]{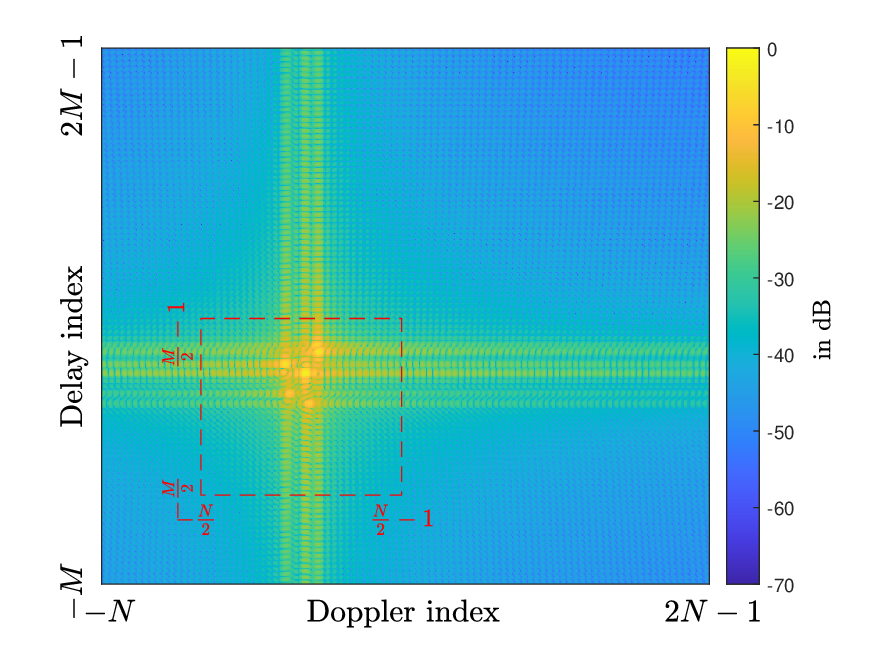}}
    \subfigure[Heatmap of $\abs{h_{\eff}}$ in $\mathcal{F}_{\exc}^{\mathrm c}$, i.e., the region which is not estimated by model-free estimation.]{\label{fig:non_mod_free_exc_pil}\includegraphics[width=0.49\linewidth]{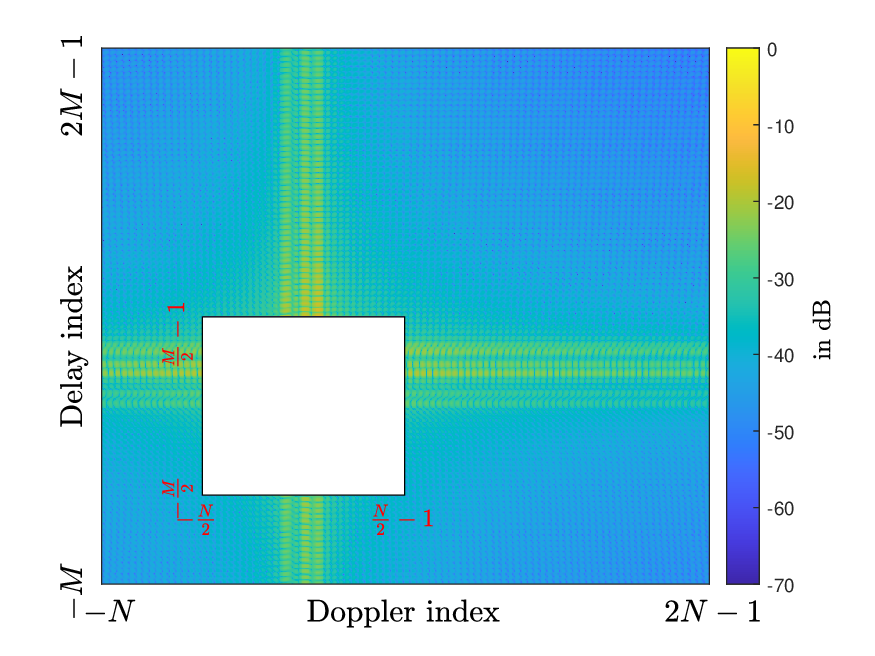}} 
    \caption{Illustration of the limitation of model-free estimation when sinc filter is used. No noise condition.}
    \label{fig:model_free_region}
\end{figure}

Before we present the proposed hybrid estimation scheme, in this subsection, we provide the rationale behind the proposed hybrid approach. 
Note that the spread of effective channel $h_{\eff}$
depends on the pulse shaping characteristics of the DD filter used. For example, when Gaussian filter is used, the resultant $h_{\eff}$ 
is well localized as shown in Fig. \ref{fig:gauss_h_eff}. However,  when a sinc filter is used, the resultant $h_{\eff}$ 
is not well localized (due to high sidelobes)
as shown in Fig. \ref{fig:sinc_h_eff}.
When model-free estimation is used with  
Gaussian filter, 
the information of $h_{\eff}$ is well captured in the model-free estimation region $\mathcal{F}_{\exc}$.
However, with a sinc filter, using model-free estimation gives only a partial estimate. 
To illustrate this, we
consider the noiseless case.
The received signal corresponding to the exclusive pilot as per \eqref{eqn:excl_pil_rsp} is shown in Fig. \ref{fig:rx_exc_pil}, which shows the quasi-periodic nature of $y_{\exc}[k',l']$. Applying the model-free estimation procedure as per \eqref{eqn:m_free_2}, we obtain the estimate $\hat{h}_{\eff}^{\free}[k,l]$ shown in Fig. \ref{fig:mod_free_exc_pil}. 
The true $h_{\eff}$ for this example as per \eqref{hdd} is shown in \ref{fig:true_h_eff_sinc}. Comparing Figs. \ref{fig:mod_free_exc_pil} and \ref{fig:true_h_eff_sinc}, 
it can be observed that the model-free estimated channel (Fig. \ref{fig:mod_free_exc_pil}) captures only a subset of the true 
effective channel (Fig. \ref{fig:true_h_eff_sinc}). That is, the effective channel outside the ${\mathcal F}_{\exc}$ region (denoted by ${\mathcal F}_{\exc}^{\mathrm c}$), shown in Fig. \ref{fig:non_mod_free_exc_pil}, is not estimated by the model-free estimation. 
This illustrates that the model-free estimate is only a partial estimate of the true effective channel, particularly when the DD filter characteristics are not well localized.  
Table~\ref{tab:quant_error} quantifies the localization difference between sinc and Gaussian filters in terms of their energy in the sidelobes and the energy outside the model-free estimation region $\mathcal{F}_{\exc}$. The energy in the sidelobes (energy outside the region $\big[-\tfrac{\tau_{\mathrm p}}{M},\tfrac{\tau_{\mathrm p}}{M}\big)\times\big[-\tfrac{\nu_{\mathrm p}}{N},\tfrac{\nu_{\mathrm p}}{N}\big)$) is $18.5\%$ and $2.35\%$ for the sinc filter and the Gaussian filter, respectively (see Fig. \ref{fig:filters} and Table~\ref{tab:quant_error}). Also, for the case of $M=N=16$ under Vehicular-A channel model, the proportion of the effective channel energy
lying outside the model-free estimation region  $\mathcal{F}_{\exc}$ (i.e., energy of $h_{\eff}$  
in $\mathcal{F}_{\exc}^{\mathrm c}$) for the sinc filter is $1.34\%$, whereas for the Gaussian filter it is almost zero ($\ll 10^{-15}\%$).

\begin{table}
\centering
\caption{Comparison of sidelobe energy and effective channel leakage beyond $\mathcal{F}_{\exc}$ for sinc and Gaussian filters.}
\begin{tabular}{|c|c|c|}
\hline
\textbf{Filter} & \makecell{\textbf{Energy in the sidelobes} 
} &\makecell{ \textbf{Energy of $h_{\eff}$ in $\mathcal{F}_{\exc}^{\mathrm{c}}$} 
}
\\
\hline
Sinc & $18.5\%$ & $1.34\% $
\\
\hline
Gaussian & $2.35\%$ & $\ll 10^{-15}\%$
\\ \hline      \end{tabular}
\label{tab:quant_error}
\end{table}

\begin{figure*}
    \centering
    \subfigure[NMSE for Gaussian and sinc filters.] 
{\label{fig:nmse_kappa}\includegraphics[width=0.34\linewidth]{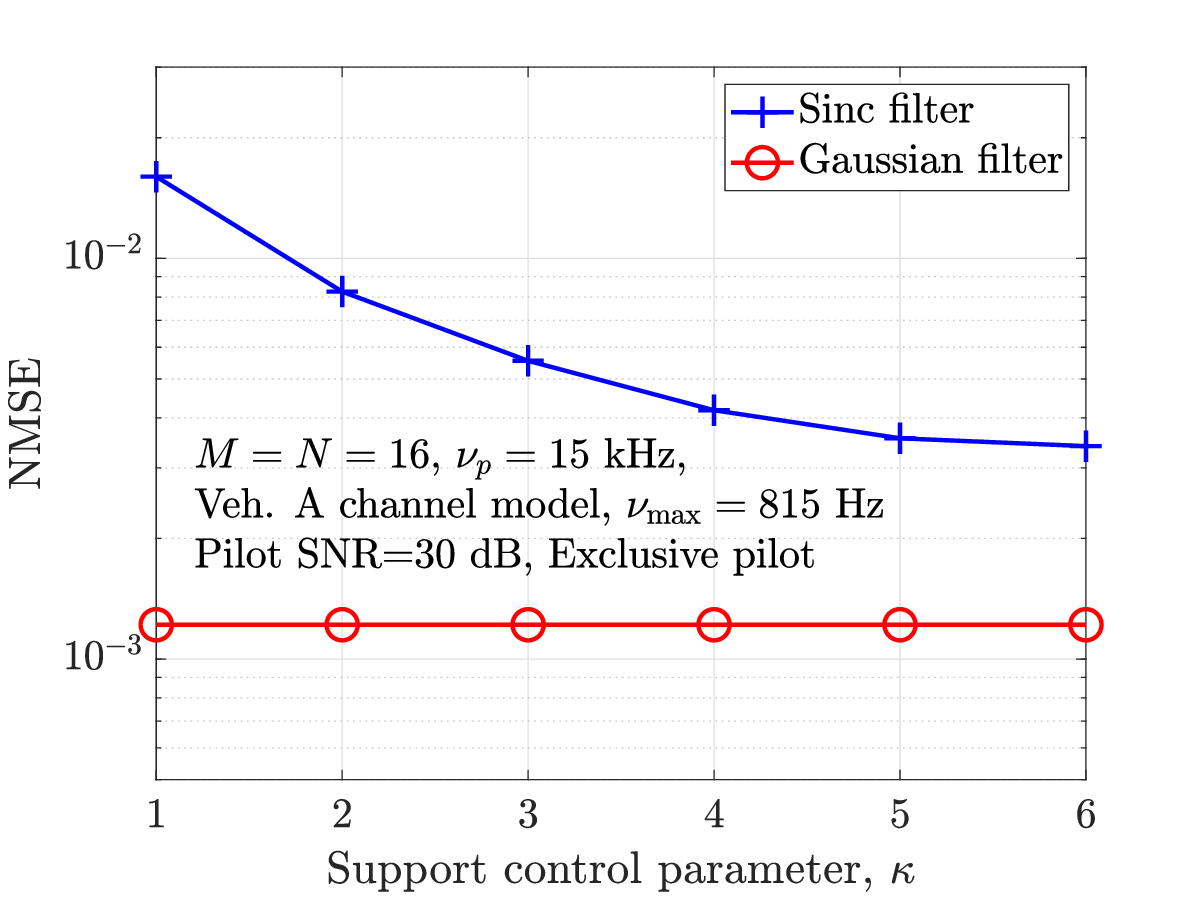}}
\hspace{-5mm}
     \subfigure[BER for Gaussian filter.]{\label{fig:ber_kappa_gauss}\includegraphics[width=0.34\linewidth]{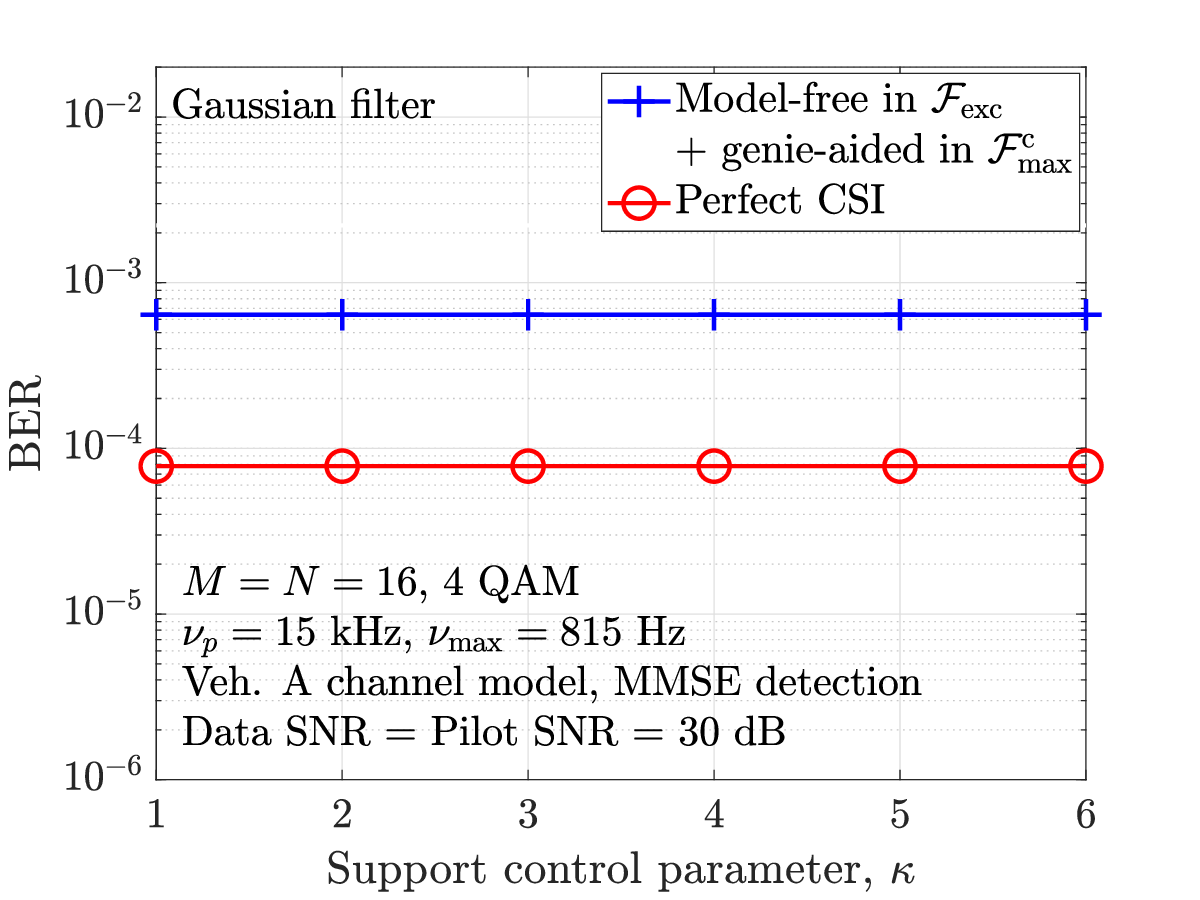}}
     \hspace{-5mm}
    \subfigure[BER for sinc filter.]{\label{fig:ber_kappa_sinc}\includegraphics[width=0.34\linewidth]{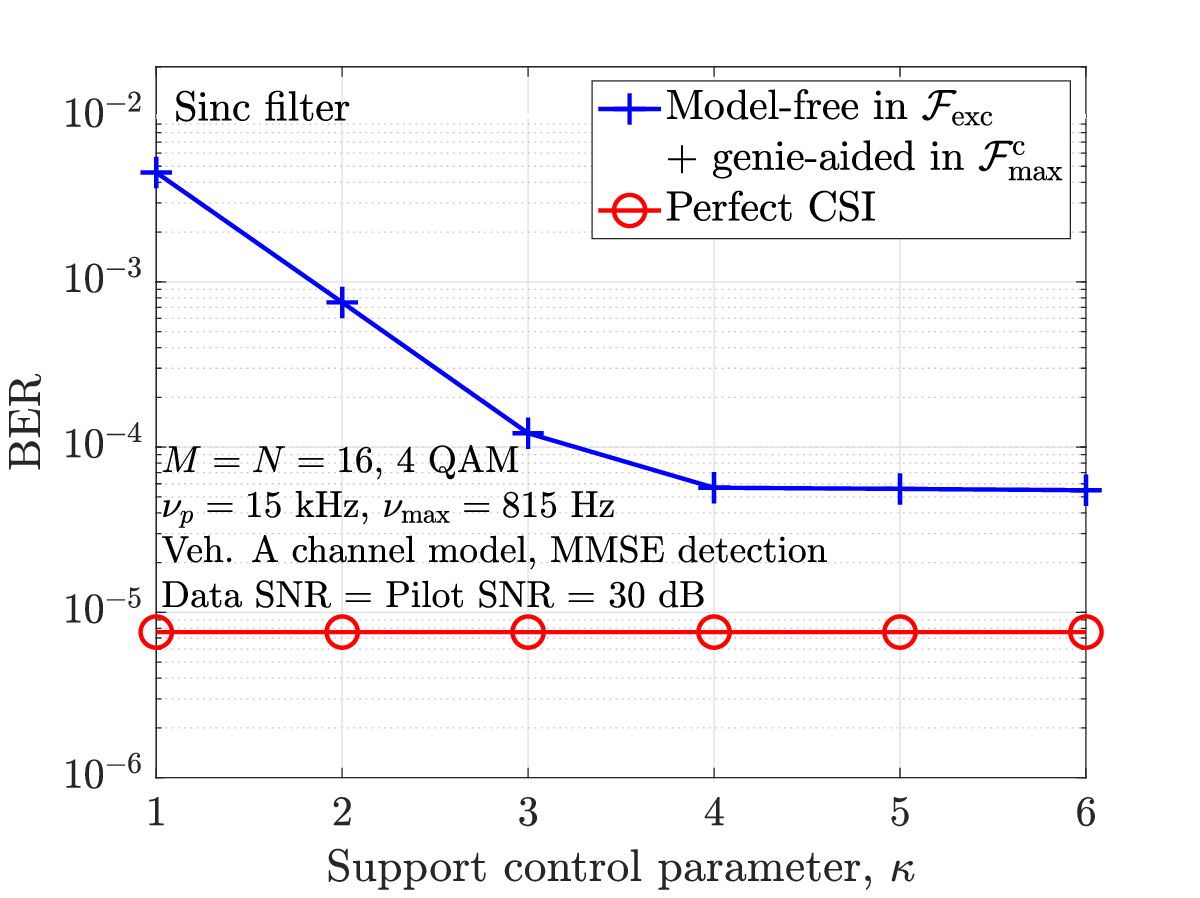}} 
    \caption{NMSE and BER performance  
    as a function of the support control parameter $\kappa$ for Gaussian and sinc filters.} 
    \label{fig:new}
\end{figure*}

We now illustrate the estimation and detection performance limitations arising from using only the partial estimate provided by model-free estimation in constructing the estimate of the $\boldH$ matrix, denoted by $\hat\boldH$. For this, in Fig. 
\ref{fig:new}, we show how the performance can improve when the effective channel outside the ${\mathcal F}_{\exc}$ region (in addition to the model-free estimates inside ${\mathcal F}_{\exc}$) is used in constructing 
$\hat\boldH$. In this illustration, we define a set ${\mathcal S}_{\kappa}$ parameterized by a positive integer $\kappa$ as 
\begin{eqnarray*}
{\mathcal S}_{\kappa} & \hspace{-2mm} = & \hspace{-2mm}
\big\{-\kappa\frac{M}{2},\cdots,\kappa\frac{M}{2}-1\big\}\hspace{-0.5mm} \times \hspace{-0.5mm} \big\{-\kappa\frac{N}{2},\cdots,\kappa\frac{N}{2}-1\big\}.  
\end{eqnarray*}
In constructing the $\hat\boldH$ matrix, 
for a given $\kappa \in {\mathbb Z}^{+}$ and $(a,b) \in {\mathbb Z}\times {\mathbb Z}$, the values of $h_{\eff}$ 
in \eqref{eqn:H_vec} are taken to be
\begin{itemize}
\item model-free estimates 
if $(a,b) \in {\mathcal F}_{\exc}$  
\item  true effective channel
if $(a,b) \in {\mathcal S}_{\kappa}\backslash{\mathcal F}_{\exc}$  
\item zero, otherwise.
\end{itemize}
Note that for $\kappa=1$, ${\mathcal S}_{\kappa}={\mathcal F}_{\exc}$, which corresponds to model-free estimation.
By varying $\kappa$ beyond 1, we control the extent of the DD support of the channel beyond ${\mathcal F}_{\exc}$ included in the construction of $\hat\boldH$. A higher value of $\kappa$ corresponds to a broader support, which incorporates more information of $h_{\eff}$. Figure \ref{fig:nmse_kappa} shows the performance in terms of NMSE, defined as the average of $\frac{||\bf{H}-\hat{\bf H}||_{F}^{2}}{||\bf{H}||_{F}^{2}}$,
as a function of $\kappa$ for Gaussian and sinc filters. Since Gaussian filter is very well localized (see Fig. \ref{fig:gauss_h_eff}), it achieves very good NMSE performance for $\kappa=1$ itself, and increasing $\kappa$ beyond 1 does not further improve the NMSE. Consequently, its BER performance also remains almost invariant to $\kappa$, as can be seen in Fig. \ref{fig:ber_kappa_gauss}. On the other hand, sinc filter's localization is not as good (see Fig. \ref{fig:sinc_h_eff}), and hence increasing $\kappa$ significantly improves the NMSE performance. A corresponding improvement in BER is observed in Fig. \ref{fig:ber_kappa_sinc}.  
This observation underscores the inadequacy of relying solely on model-free estimation, particularly in scenarios where the effective channel is not well localized. Also, comparing the
perfect channel state information (CSI) based BER performance of Gaussian and sinc filters in Figs. \ref{fig:ber_kappa_gauss} and \ref{fig:ber_kappa_sinc}, respectively, we see that the sinc filter achieves a much better BER ($8\times 10^{-6}$) compared to that of the Gaussian filter ($8\times 10^{-5}$). This is because, while the sinc filter has nulls at the information grid points (i.e., no inter-symbol interference (ISI) at the DD sampling points), the Gaussian filter has non-zero values in the main lobe at the sampling points (causing ISI). This indicates the potential for achieving better BER performance using a sinc filter compared to using a Gaussian filter, and this, however, requires the estimation performance to be improved beyond that of the naive model-free estimation. Towards accomplishing this,
we devise a novel estimation approach in the following subsection.

\subsubsection{Proposed hybrid estimation for exclusive pilot}
In the proposed approach, we retain the model-free estimate of the channel in ${\mathcal F}_{\exc}$ (for simplicity in acquiring fractional DD) and obtain a low-complexity model-dependent estimate of the channel in ${\mathcal F}_{\exc}^{\mathrm c}$ (for  performance enhancement). In model-dependent estimation, an estimate of the physical channel parameters $\{\tau_i,\nu_i,h_i\}, i=1,\ldots,P$, is obtained. A primitive model-dependent estimation approach based on energy detection is used in \cite{bits2}. In this method, the DD grid points where the received signal has high-energy peaks are interpreted as potential channel paths, and the corresponding path gains are estimated using a least squares (LS) approach. This method is effective when the channel exhibits integer DD shifts (see Fig. 11 in \cite{bits2} which considers integer DD channel). However, in practice, the channel spread often consists of fractional DD components, where this method suffers due to the approximation of fractional shifts to the nearest integer grid points, leading to performance degradation (see Fig. 10 in \cite{bits2} which considers Veh-A channel).
To overcome this limitation, we modify the energy-based approach to better suit fractional DD scenarios and integrate it into our hybrid estimation framework. The hybrid estimate \( \hat{h}_{\text{eff}}^{\text{hyb}} \) is constructed by combining the model-free estimate \( \hat{h}_{\text{eff}}^{\text{free}} \) with the model-dependent estimate \( \hat{h}_{\text{eff}}^{\text{dep}} \), where \(\hat{h}_{\eff}^{\dep}\) is the effective channel obtained from substituting the model-dependent channel parameter estimates in \eqref{hdd}. The values of \( \hat{h}_{\text{eff}}^{\free} \) provided by the model-free estimate are retained as is in \(\mathcal{F}_{\exc}\), while the remaining entries are filled using \(\hat{h}_{\eff}^{\dep}\).

\begin{figure*}
    \centering
    \subfigure[Heatmap of $|\hat{h}_{\eff}^{\free}|$ obtained through model-free estimate.]{\label{fig:m_free_demo}\includegraphics[width=0.3\linewidth]{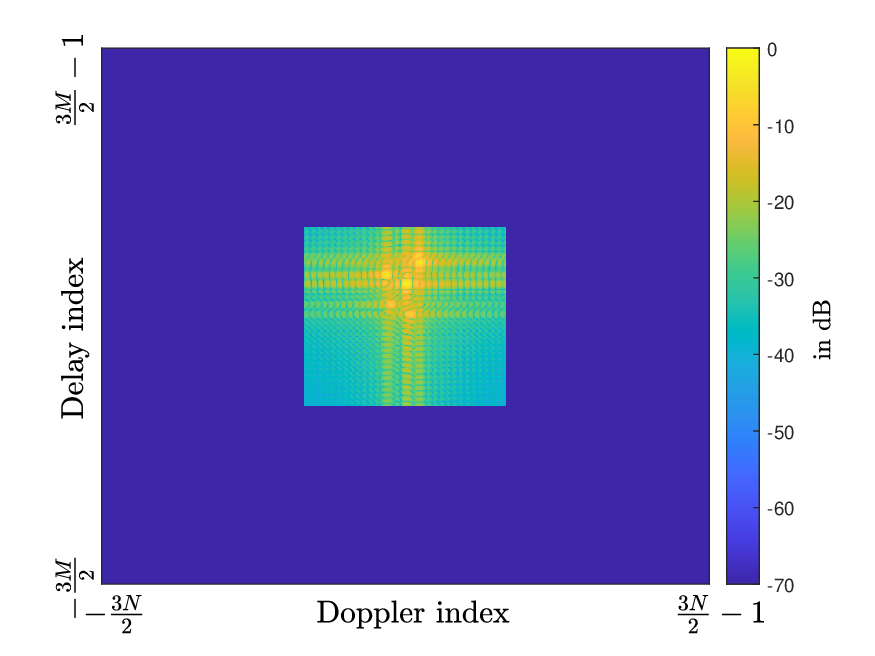}}
    \raisebox{0.1\linewidth}{+}
      \subfigure[ Heatmap of $|\hat{h}_{\eff}^{\dep}|$ obtained through model-dependent estimate in the \(\mathcal{F}_{\exc}^c\) region, using the proposed \textbf{Algorithm}~\ref{alg:mdep_est}.]{\label{fig:m_dep_demo}\includegraphics[width=0.3\linewidth]{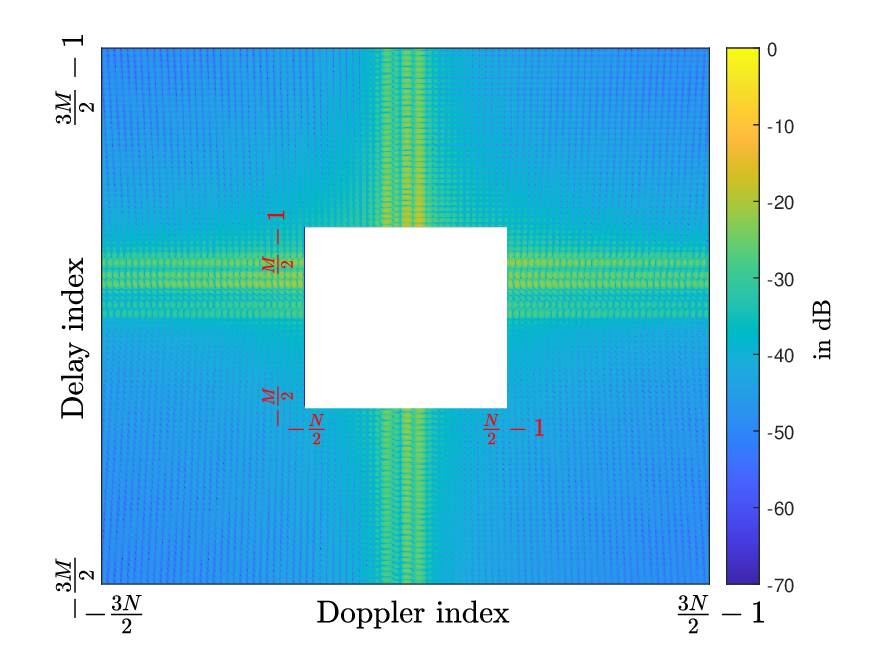}}
      \raisebox{0.1\linewidth}{=}
     \subfigure[Heatmap of $|\hat{h}_{\eff}^{\hyb}|$ obtained by combining mode-free and model-dependent estimates.]{\label{fig:hyb_demo}\includegraphics[width=0.3\linewidth]{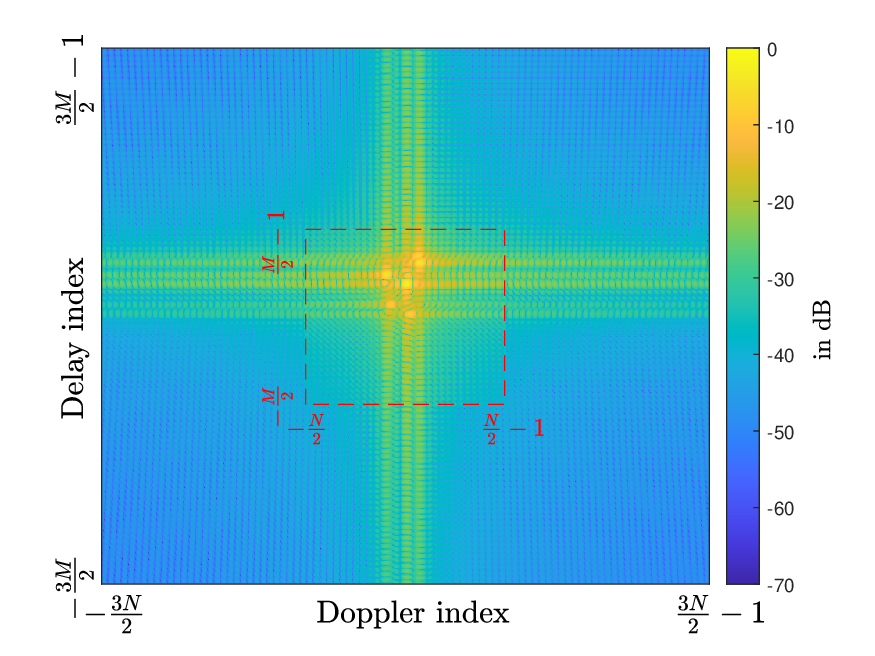}}
    \caption{Illustration of obtaining $\hat{h}_{\eff}^{\hyb}$ from $\hat{h}_{\eff}^{\free}$ and  $\hat{h}_{\eff}^{\dep}$ with exclusive pilot frame.} 
    \label{fig:hyb_filling}
\end{figure*}
\subsubsection{Proposed low-complexity model-dependent algorithm} 
We define a binary mask matrix \( mask \in \{0,1\}^{M \times N} \), where the entries surrounding the exclusive pilot location are set to one and the rest to zero, as
\begin{equation}
mask[k,l] = \begin{cases}
1, & \text{if } (k,l) \in \mathcal{K} \times \mathcal{L}, \\
0, & \text{otherwise},
\end{cases}
\label{eqn:mask}
\end{equation}
where $\mathcal{K} = \left\{ \tfrac{M}{2}, \dotsc, \tfrac{M}{2} + k_{\max} \right\}$, $\mathcal{L} = \left\{ \tfrac{N}{2} - l_{\max}, \dotsc, \tfrac{N}{2} + l_{\max} \right\}$, 
\( k_{\max} = \left\lceil \tfrac{\tau_{\max}}{\tau_p/M} \right\rceil \),  
\( l_{\max} = \left\lceil \tfrac{\nu_{\max}}{\nu_p/N} \right\rceil \).
The received pilot frame is masked using this window as
\[
y_{\text{exc}}^{\text{mask}}[k,l] = y_{\text{exc}}[k,l] \cdot mask[k,l], 
\]
$\forall k \in \{0,\dotsc,M-1\},~ l \in \{0,\dotsc,N-1\}$. 

\textbf{Estimation of path delays and Dopplers:}
To estimate the DD paths, we adopt an energy-based detection approach due to its low complexity. However, practical channels often exhibit fractional DDs, which cannot be captured accurately by pure energy-based estimators restricted to the integer DD grid. To address this, we include a simple refinement step to approximate fractional paths using adjacent bins.

The estimation proceeds iteratively. In each iteration, the DD bin with the maximum energy in the masked received signal \(y_{\exc}^{\mask}\) is identified as an integer DD path. We then examine the adjacent bins\footnote{For a bin \((a, b) \in \mathbb{Z} \times \mathbb{Z}\), we define the adjacent bins
as the bins in locations \(\{a-1, a, a+1\} \times \{b-1, b, b+1\} \setminus \{(a, b)\}\).}  
of this location. The adjacent bin with the highest energy is selected, and the average of its location with that of the previously identified bin is used to obtain a second (fractional) DD path.
This method allows us to account for fractional DD paths in an approximate sense, without invoking high-complexity techniques\footnote{More accurate estimators exist for resolving fractional paths, but they are computationally intensive and would defeat the simplicity of the model-free baseline. Our goal is not to recover the exact fractional values, but to introduce a low-complex refinement that meaningfully enhances the estimate.}.
These steps are formalized in \textbf{Algorithm}~\ref{alg:mdep_est}.

In iteration \(t\), the DD bin with the highest energy is denoted by \((p_1, q_1)\) (step 4 of \textbf{Algorithm}~\ref{alg:mdep_est}). A bin \((p_2, q_2)\) is selected from the adjacent bins as the one with the maximum energy (step 6 in \textbf{Algorithm}~\ref{alg:mdep_est}). If all adjacent bins have zero energy, we randomly pick one of the adjacent bins\footnote{
To assess the effect of this random bin selection, we ran simulations for three cases, namely, 1) random adjacent bin selection, 2) selecting the adjacent bin nearest to the pilot, and 3) selecting the adjacent bin farthest from the pilot. Simulation results showed that the performance of all the three cases are roughly the same. This is because the situation where all adjacent bins of a detected high-energy bin become zero arises only after several paths have already been estimated, and the additional path chosen in such a case does not have sufficient strength to significantly affect the overall performance, leading to roughly the same performance for all the three cases.}.
The estimated DD pairs for iteration \(t\) are given by
\begin{equation*}
{\left(\!\left(p_1\! -\! \tfrac{M}{2}\right) \!\tfrac{\tau_p}{M},\! \left(q_1\! -\! \tfrac{N}{2}\right) \!\tfrac{\nu_p}{N} \right)\!,\!\left(\!\left(\tfrac{p_1 + p_2}{2} \!- \!\tfrac{M}{2}\right)\! \tfrac{\tau_p}{M}, \hspace{-0.5mm} \left(\tfrac{q_1 + q_2}{2} \!- \!\tfrac{N}{2}\right) \!\tfrac{\nu_p}{N}\! \right)}
\end{equation*}
corresponding to the integer and fractional paths, respectively. After each iteration, the entries at \((p_1, q_1)\) and \((p_2, q_2)\) in \( y_{\text{exc}}^{\text{mask}} \) are zeroed out so that they are not picked again. The process continues until all entries in \(y_{\exc}^{\mask}\) are exhausted. Since at least one bin is made zero in every iteration, the maximum number of iterations in this algorithm is of the order \( \mathcal{O}(k_{\max} l_{\max}) \).

{\bf Estimation of path gains:}
Having obtained the estimated path DD parameters \( \hat{\boldsymbol{\tau}} \) and \( \hat{\boldsymbol{\nu}} \), 
the corresponding path gains are computed as follows. Towards this, an equivalent system model of \eqref{eqn:sys_mod_vec} can be written as
\begin{align}
\boldy =\underbrace{[\boldsymbol{\phi}_1~ \boldsymbol{\phi}_2~ \cdots~ \boldsymbol{\phi}_P]}_{\triangleq \boldsymbol{\Phi}(\boldtau,\boldnu)} \boldh + \boldn,
\label{eqn:alter_sys_model}
\end{align}
where \( \boldsymbol{\phi}_i = \boldH(\tau_i, \nu_i, 1) \boldx \), and \(\boldh = [h_1~ h_2~ \cdots~ h_P]^{T} \). $\boldsymbol{\phi}_i$ can be interpreted as the received signal when the channel has only the $i$th path with its corresponding path gain as $1$. Using this alternate system model, the LS estimate of the path gains is given by
\begin{equation}
\hat{\mathbf{h}} = \left[ \boldsymbol{\Phi}^{H}(\hat{\boldsymbol{\tau}}, \hat{\boldsymbol{\nu}}) \boldsymbol{\Phi}(\hat{\boldsymbol{\tau}}, \hat{\boldsymbol{\nu}}) \right]^{-1} \boldsymbol{\Phi}^{H}(\hat{\boldsymbol{\tau}}, \hat{\boldsymbol{\nu}}) \boldy_{\exc},
\label{eqn:h_hat}
\end{equation}
where $\boldsymbol{\Phi}(\hat{\boldtau},\hat{\boldnu})=[\boldH(\hat{\tau}_1,\hat{\nu}_1,1)\boldx_{\exc}~\boldH(\hat{\tau}_2,\hat{\nu}_2,1)\boldx_{\exc}~\cdots]$ is constructed 
using $(\hat{\boldtau},\hat{\boldnu})$
from {\bf Algorithm}~\ref{alg:mdep_est} and the exclusive pilot vector ${\bf x}_{\exc}$.

\begin{algorithm}[t]
\caption{{\normalsize Proposed low-complexity algorithm for estimation of path 
delays and Dopplers}}
\label{alg:mdep_est}
    \begin{algorithmic}[1]
        \STATE \textbf{Input: }Masked pilot frame $y_{\exc}^{\mask}[k,l]=y_{\exc}[k,l]\cdot\mask[k,l],\mathcal{E}[k,l]=\abs{y_{\exc}^{\mask}[k,l]}^2 ~\forall~k\in\{0,\cdots,M-1\},l\in\{0,\cdots,N-1\}$
        \STATE \textbf{Initialization: }
        $\hat{\boldtau}=[~], \hat{\boldnu}=[~]$, $t=1$
        \STATE \textbf{repeat}
        \STATE \hspace{5mm}
        $[p_1,q_1]=\underset{\substack{r\in\{0,1,\cdots,M-1\}\\ s\in\{0,1,\cdots,N-1\}}}{\text{arg max}}\mathcal{E}[r,s]$
        \STATE \hspace{5mm} $\hat{\boldtau}=[\hat{\boldtau},(p_1-\frac{M}{2})\frac{\taup}{M}],\hat{\boldnu}=[\hat{\boldnu},(q_1-\frac{N}{2})\frac{\nup}{N}]$
        \STATE \hspace{5mm}$[p_2,q_2]=\underset{\substack{(r,s)\in\{p_1-1,p_1,p_1+1\}\times\\\{q_1-1,q_1,q_1+1\}\backslash\{(p_1,q_1)\}}}{\text{arg max}}\hspace{-5mm}\mathcal{E}[r,s]$
        \STATE\hspace{5mm} $\hat{\boldtau}=[\hat{\boldtau}, \left(\frac{p_1+p_2}{2}-\frac{M}{2}\right)\frac{\taup}{M}],\hat{\boldnu}=[\hat{\boldnu},\left(\frac{q_1+q_2}{2}-\frac{N}{2}\right)\frac{\nup}{N}]$
        \STATE \hspace{5mm} $\mathcal{E}[p_1,q_1]=0,\mathcal{E}[p_2,q_2]=0$
        \STATE \hspace{5mm} $t=t+1$
        \STATE\textbf{until }$\mathcal{E}[k,l]=0~\forall~k\in\{0,\cdots,M-1\},l\in\{0,\cdots,N-1\}$
        \STATE\textbf{Output: }Estimated parameters $\hat{\boldtau},\hat{\boldnu}$.
    \end{algorithmic}
\end{algorithm}

\textbf{Hybrid construction of the effective channel:}  
The model-free estimate \( \hat{h}_{\text{eff}}^{\text{free}}[a,b] \) is available over the region \( \mathcal{F}_{\exc} \), and the model-dependent estimate \( \hat{h}_{\text{eff}}^{\text{dep}}[a,b] \), obtained using \textbf{Algorithm}~\ref{alg:mdep_est}, provides estimates over the entire DD space. These two estimates are combined to form a hybrid estimate as follows:
\begin{equation}
\hat{h}_{\text{eff}}^{\text{hyb}}[a,b] =
\begin{cases}
\hat{h}_{\text{eff}}^{\text{free}}[a,b], & \text{if}~ (a,b)\in\mathcal{F}_{\exc}, \\
\hat{h}_{\text{eff}}^{\text{dep}}[a,b], & \text{otherwise}.
\end{cases}
\label{eqn:h_eff_hyb}
\end{equation}
Figure~\ref{fig:hyb_filling} illustrates this hybrid construction. Figure~\ref{fig:m_free_demo} shows the model-free estimate, limited to the region \( \mathcal{F}_{\exc} \). Figure~\ref{fig:m_dep_demo} presents the model-dependent estimate, obtained based on the estimated channel path parameters using \textbf{Algorithm}~\ref{alg:mdep_est}. The hybrid estimate, shown in Fig.~\ref{fig:hyb_demo}, uses model-free values within \( \mathcal{F}_{\exc} \) and model-dependent values outside it. This hybrid estimate \( \hat{h}_{\text{eff}}^{\text{hyb}} \) is then used to construct the estimated effective channel matrix \( \hat{\mathbf{H}} \) using \eqref{eqn:H_vec} for data detection.

\subsection{Estimation for embedded pilot frame}
The exclusive pilot frame design, while providing the advantage of interference-free channel estimation due to complete separation between pilot and data, suffers from reduced throughput. To alleviate this throughput penalty, we consider an embedded pilot frame structure in which both pilot and data symbols coexist in the same frame.
In the embedded frame structure, in addition to the centrally placed pilot symbol \( \alpha \), data symbols occupy the rest of the frame except for a defined guard region around the pilot to reduce interference between data and pilot (see Fig.~\ref{fig:emb_pil}). 
The embedded pilot frame is  
constructed as
\begin{equation}
    x_{\text{emb}}[k,l] =
    \begin{cases}
        \alpha, & \text{if } k = \tfrac{M}{2},~ l = \tfrac{N}{2}, \\
        0, & (k,l) \in \mathcal{G}, \\
        \text{data}, & \text{otherwise},
    \end{cases}
\end{equation}
where the data symbols are drawn from the modulation alphabet \( \mathbb{A} \), and \( \mathcal{G} \) denotes the guard region defined to mitigate pilot-data interference. The guard region \( \mathcal{G} \) is specified as
\begin{equation}
\mathcal{G} = \mathcal{G}' \setminus \left( \tfrac{M}{2}, \tfrac{N}{2} \right),
\end{equation}
where
\begin{equation}
\mathcal{G}'\! =\! \left\{ \tfrac{M}{2}\! -\! 4k_{\max}, \cdots, \tfrac{M}{2} + 4k_{\max} \right\}\! \times\! \left\{ 0, \cdots, N-1 \right\}.
\end{equation}

\subsubsection{Model-free estimation for embedded frame}
For model-free estimation in the embedded frame, we read-off a subset of the guard region to avoid contamination from data symbols. This read-off region, denoted by \( \mathcal{R}_{\text{emb}} \), is defined as
\begin{equation}
\mathcal{R}_{\text{emb}}\! =\! \left\{ \tfrac{M}{2} - 2k_{\max}, \dotsc, \tfrac{M}{2} + 2k_{\max} \right\} \times \left\{ 0, 1, \dotsc, N-1 \right\}.
\end{equation}
The corresponding model-free estimation region \(\mathcal{F}_{\emb}\) is 
\begin{equation}
    \mathcal{F}_{\emb}=\{-2k_{\max},\cdots,2k_{\max}\}\times\{-\tfrac{N}{2},\cdots,\tfrac{N}{2}-1\}.
\end{equation}
The model-free estimate of the effective channel for embedded pilot frame, denoted by $\hat{h}_{\text{eff}}^{\fremb}$, is computed as
\begin{equation}
\hat{h}_{\text{eff}}^{\fremb}[a,b] \hspace{-0.5mm} = \hspace{-0.5mm}
\begin{cases}
y_{\text{emb}}[a\!+\!\tfrac{M}{2}, b\!+\!\tfrac{N}{2}]\dfrac{e^{-j\pi a/N}}{\alpha}, & \hspace{-2mm} \text{if } (a,b)\! \in\! \mathcal{F}_{\text{emb}}, \\
0, & \hspace{-2mm} \text{otherwise},
\end{cases}
\end{equation}
where \( y_{\text{emb}} \) is the received DD signal of the embedded pilot frame. Note that the size of embedded model-free estimation region \( \mathcal{F}_{\text{emb}} \) is smaller than the exclusive model-free estimation region \(\mathcal{F}_{\exc}\), leading to an even more partial estimate in this case.

\subsubsection{Hybrid estimation for embedded frame}
To overcome the limitation of partial estimation, we extend the hybrid estimation strategy to embedded pilot frames. Specifically, we apply the proposed model-dependent estimation
using the masking approach. When applied to the embedded frame, the $mask$ in \eqref{eqn:mask} has non-zero entries only over \( \mathcal{R}_{\text{emb}} \), and this isolates the region required for estimation.
The masked signal is
\begin{equation}
y_{\text{emb}}^{\text{mask}}[k,l] = y_{\text{emb}}[k,l] \cdot mask[k,l].
\end{equation}
Assuming that the interference from data to pilot is negligible, the above masked signal can now be directly used in \textbf{Algorithm}~\ref{alg:mdep_est} to estimate the DD parameters and corresponding path gains without requiring any algorithmic changes. The hybrid estimate of the effective channel is then constructed as
\begin{equation}
\hat{h}_{\text{eff}}^{\hybemb}[a,b] =
\begin{cases}
\hat{h}_{\text{eff}}^{\fremb}[a,b], & \text{if } (a,b) \in \mathcal{F}_{\text{emb}}, \\
\hat{h}_{\text{eff}}^{\depemb}[a,b], & \text{otherwise},
\end{cases}
\end{equation}
where \(\hat{h}_{\eff}^{\depemb}\) is the effective channel obtained from the model-dependent DD estimates of the embedded frame using \eqref{hdd}. Using \(\hat{h}_{\eff}^{\hybemb}\), the  \(\hat{\boldH}\) matrix is constructed using \eqref{eqn:H_vec}, and detection of data symbols proceeds using  
\(\hat{\boldH}\).

\textit{Complexity:}
The additional computational complexity introduced by the hybrid estimation arises from estimating the paths and constructing \( h_{\text{eff}}^{\text{dep}} \) over regions not covered by model-free estimation. The maximum number of paths that can be estimated, denoted by $\hat{P}_{\max}$, is 
\( \hat{P}_{\max} = 2(k_{\max}+1)(2l_{\max}+1) \). The total computational burden is summarized in Table~\ref{complexity}, where the per-operation costs are listed. In Table~\ref{complexity}, \( \beta = (6M-1)(6N-1) - MN \) denotes the total number of points where \( h_{\text{eff}}^{\text{dep}}(\tau, \nu) \) must be evaluated. The hybrid estimation method, when applied to the embedded pilot frame, strikes a favorable balance between estimation accuracy and throughput efficiency, leveraging structured pilot-data coexistence without incurring significant additional complexity.

\begin{table}
\caption{Additional computational complexity for hybrid estimation}
\resizebox{\linewidth}{!}{
\begin{tabular}{|c|c|c|c|}
\hline
\textbf{Operation} & \textbf{No. of real multiplications} & \textbf{No. of real additions} & \textbf{Total complexity} \\ \hline
\( \Phi^H\Phi \) & \( 4\hat{P}_{\max}^2MN \) & \( 2\hat{P}_{\max}^2(2MN - 1) \) & \( 8\hat{P}_{\max}^2MN - 2\hat{P}_{\max}^2 \) \\ \hline
\( (\Phi^H\Phi)^{-1} \) & -- & -- & \( \mathcal{O}(\hat{P}_{\max}^3) \) \\ \hline
\( \Phi^H\boldy \) & \(4\hat{P}_{\max}MN \) & \( 2\hat{P}_{\max}(2MN - 1) \) & \( 8\hat{P}_{\max}MN - 2\hat{P}_{\max} \) \\ \hline
\( (\Phi^H\Phi)^{-1}\Phi^H\boldy \) & \( 4\hat{P}_{\max}^2 \) & \( 4\hat{P}_{\max}^2 - 2\hat{P}_{\max} \) & \( 8\hat{P}_{\max}^2 - 2\hat{P}_{\max} \) \\ \hline
\( h_{\text{eff}}^{\text{dep}}(\tau, \nu) \) & \( 20\hat{P}_{\max}\beta \) & \( 2(5\hat{P}_{\max} - 1)\beta \) & \( (30\hat{P}_{\max} - 2)\beta \) \\ \hline
\end{tabular}}
\label{complexity}
\end{table}

\section{Results and Discussions} 
\label{results}
In this section, we present the evaluated performance of the proposed hybrid estimation scheme for both exclusive and embedded pilot frame configurations. The frame size parameters ($M$ and $N$ values) are related to the bandwidth and time duration of transmission and the fundamental DD periods as  
$B=M\nup$ and $T=N\taup$.
The parameters for the simulation experiments are chosen as follows. 
The Doppler period is set to \(\nu_{\mathrm p} = 15\) kHz and the delay period is, therefore, \(\tau_{\mathrm p} = 1/\nu_{\mathrm p} = 66.67~\mu\)s. The channel is assumed to have \(P=6\) paths with path delays and powers specified according to the power delay profile (PDP) of the Vehicular-A channel model (see Table~\ref{tab:veh_A}) \cite{vehA}, where the maximum delay spread is 2.51 $\mu$s.
In terms of Doppler, the Doppler shift of the \(i\)th path is modeled as \(\nu_{\max}\cos(\theta_i)\), where \(\theta_i, i=1,2,\cdots,P,\) is drawn independently and uniformly from \([0,2\pi)\), and \(\nu_{\max}\) is taken to be 815 Hz\footnote{This choice of \(\nu_{\max}=815\) Hz corresponds to a carrier frequency (\(f_c\)) of \(4\) GHz and a maximum velocity of \(220\) km/h, since \(\nu_{\max}=\frac{f_c v}{c}=\frac{4\times 10^9 \times (220 \times 10^3/3600)}{3\times 10^8} \approx 815~\text{Hz}\).} \cite{bits2}. This \(\nu_{\max}\) of 815 Hz corresponds to a maximum Doppler spread of \(2\nu_{\max}=1.63\) kHz. For all the performance figures other than Fig.~\ref{fig:TDL}, the Vehicular-A channel model as described above is used. For Fig.~\ref{fig:TDL}, we have used the 3GPP-specified TDL-A and TDL-C channel models \cite{3gpp} with a larger number of paths ($P=23,24$) and a larger Doppler $(\nu_\mathrm{max}=3$ kHz).
The frame size parameters (number of delay bins in $\tau_\mathrm{p}$ and number of Doppler bins in $\nu_\mathrm{p}$) are taken to be \(M = 64\) and \(N = 24\), respectively\footnote{The choice of $M=64$, $N=24$ corresponds to a medium/large Zak-OTFS frame size and follows the setting adopted in \cite{bits2}. We have also carried out simulations for a system with a smaller frame size of $M=N=16$ and observed similar performance gains for the proposed scheme observed for $M=64$, $N=24$. To avoid repetition, we present results only for $M=64$, $N=24$.}, which correspond to a bandwidth of $B=M\nup=64\times 15$ kHz $=960$ kHz and a frame duration of $T=N\taup=24\times66.67~\mu$s $=1.6$ ms.
For the considered parameters, \(k_{\max}=3\) and \(l_{\max}=2\). The path powers are normalized such that \(\sum_{i=1}^P \mathbb{E}[|h_i|^2]=1\).
For embedded pilot frames, the pilot-to-data power ratio (PDR) is defined as \(E_{\mathrm p}/E_{\mathrm d}\), where \(E_{\mathrm p}\) and \(E_{\mathrm d}\) denote pilot and data energies, respectively. The data SNR (DSNR) is given by \(E_{\mathrm d}/(N_0MN)\) and the pilot SNR (PSNR) is given by \(E_{\mathrm p}/(N_0MN)\).
For the chosen Vehicular-A channel parameters, the guard region in the embedded frame occupies \(37.5\%\) of the bins, yielding an overall frame efficiency of \(62.5\%\). Detection is carried out using an LMMSE equalizer, followed by minimum-distance decoding.
All the simulations are run on a 12th Gen Intel(R) Core(TM) i7-12700F processor operating at 2.10 GHz using MATLAB R2024b.

\begin{table}[t]
\centering
\caption{Vehicular-A Channel PDP}
\begin{tabular}{|c|c|c|c|c|c|c|}
\hline
Path index \( i \) & 1 & 2 & 3 & 4 & 5 & 6 \\ \hline
Delay \( \tau_i \) (in \( \mu \)\text{s}) & 0 & 0.31 & 0.71 & 1.09 & 1.73 & 2.51 \\ \hline
Path power (in dB) & 0 & -1 & -9 & -10 & -15 & -20 \\ \hline
\end{tabular}
\label{tab:veh_A}
\end{table}

\begin{figure}
    \centering
    \includegraphics[width=0.875\linewidth]{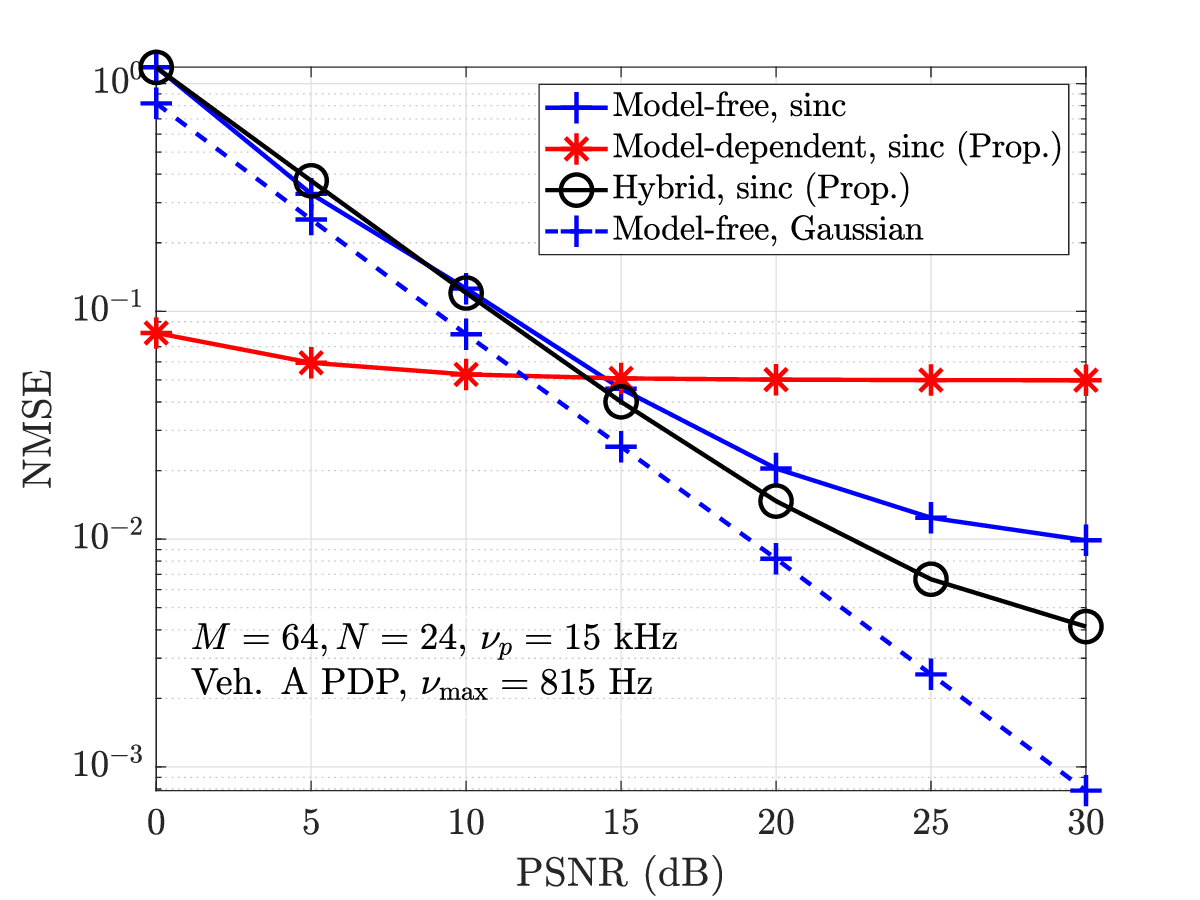}
    \caption{NMSE performance of the proposed hybrid estimation scheme as a function of PSNR for exclusive pilot frame.}
    \label{fig:nmse_exclusive_pilot}
\end{figure}

\subsection{Performance with exclusive pilot frame}
\subsubsection{Estimation accuracy}
Figure~\ref{fig:nmse_exclusive_pilot} shows the NMSE performance of the proposed hybrid estimation scheme as a function of PSNR for sinc filter with exclusive pilot frame. The individual performance of model-free estimation and the proposed low-complexity model-dependent estimation are shown for comparison. Model-free estimation performance with the Gaussian filter is also shown. The following observations can be made from Fig.~\ref{fig:nmse_exclusive_pilot}. 
At low PSNR values, the model-free estimate exhibits higher NMSE compared to that of the proposed model-dependent estimate, primarily due to the dominant influence of noise in the low PSNR regime. However, as the PSNR increases, the NMSE of the model-dependent estimate floors at a value much higher than that of the model-free estimate. Also, the proposed hybrid estimate, like the model-free estimate, underperforms relative to the model-dependent estimate at low PSNRs as a consequence of the inaccurate model-free estimate at low PSNRs, i.e., the poor quality of the model-free estimate limits the overall effectiveness of the hybrid scheme in the low PSNR regime. However, beyond a certain PSNR (around 15 dB), the quality of the model-free estimate improves significantly, and, as a result, the hybrid estimate begins to outperform both the model-free and model-dependent estimates, with the performance gap widening at higher PSNR values. Further, it can be seen that the model-free estimation using the Gaussian filter achieves superior NMSE performance compared to all those of the estimators with the sinc filter. This can be attributed to the excellent localization properties of the Gaussian filter in the DD domain.

\begin{figure}
    \centering
    \includegraphics[width=0.875\linewidth]{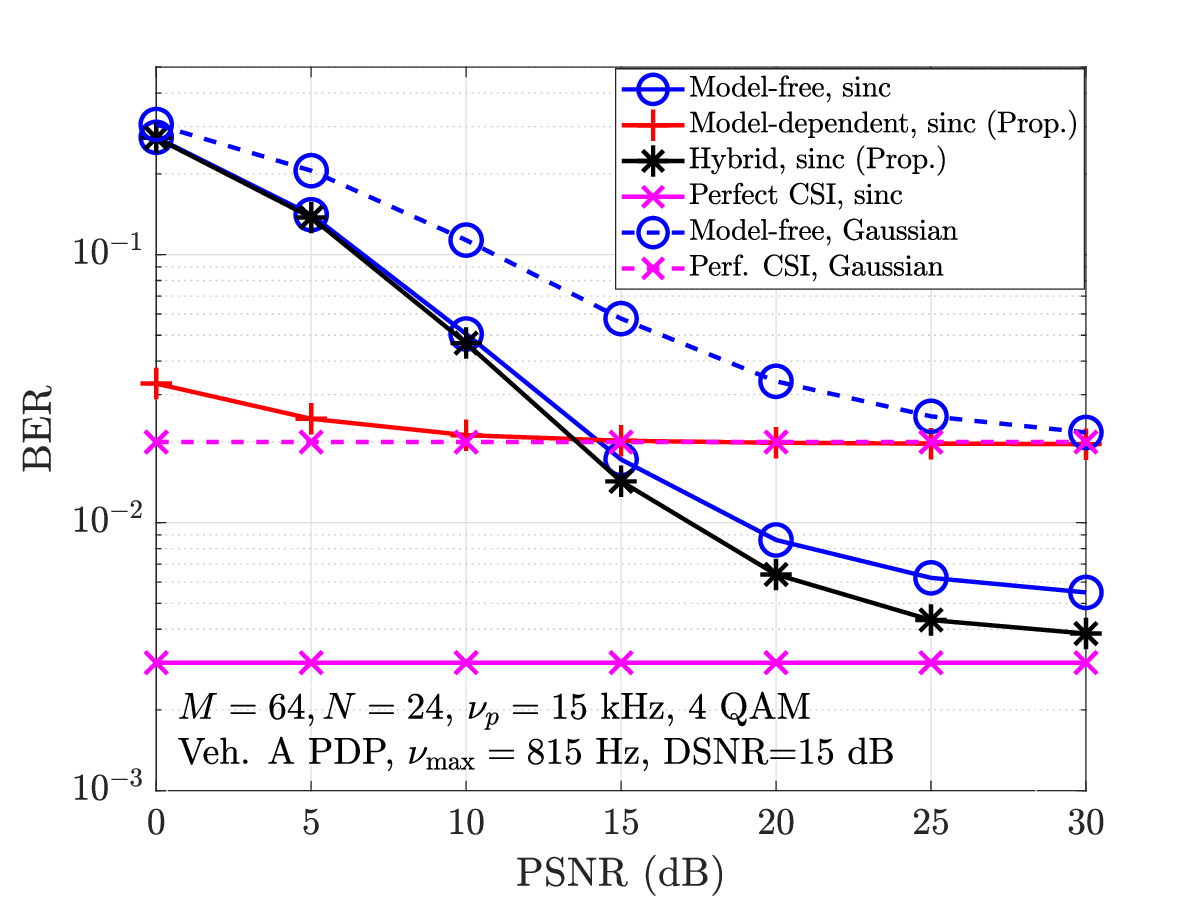}
    \caption{BER performance of the proposed hybrid estimation scheme as a function of PSNR for exclusive pilot frame at 15 dB DSNR.}
    \label{fig:ber_vs_psnr_exclusive_pilot}
\end{figure}
\begin{figure}
    \centering
    \includegraphics[width=0.875\linewidth]{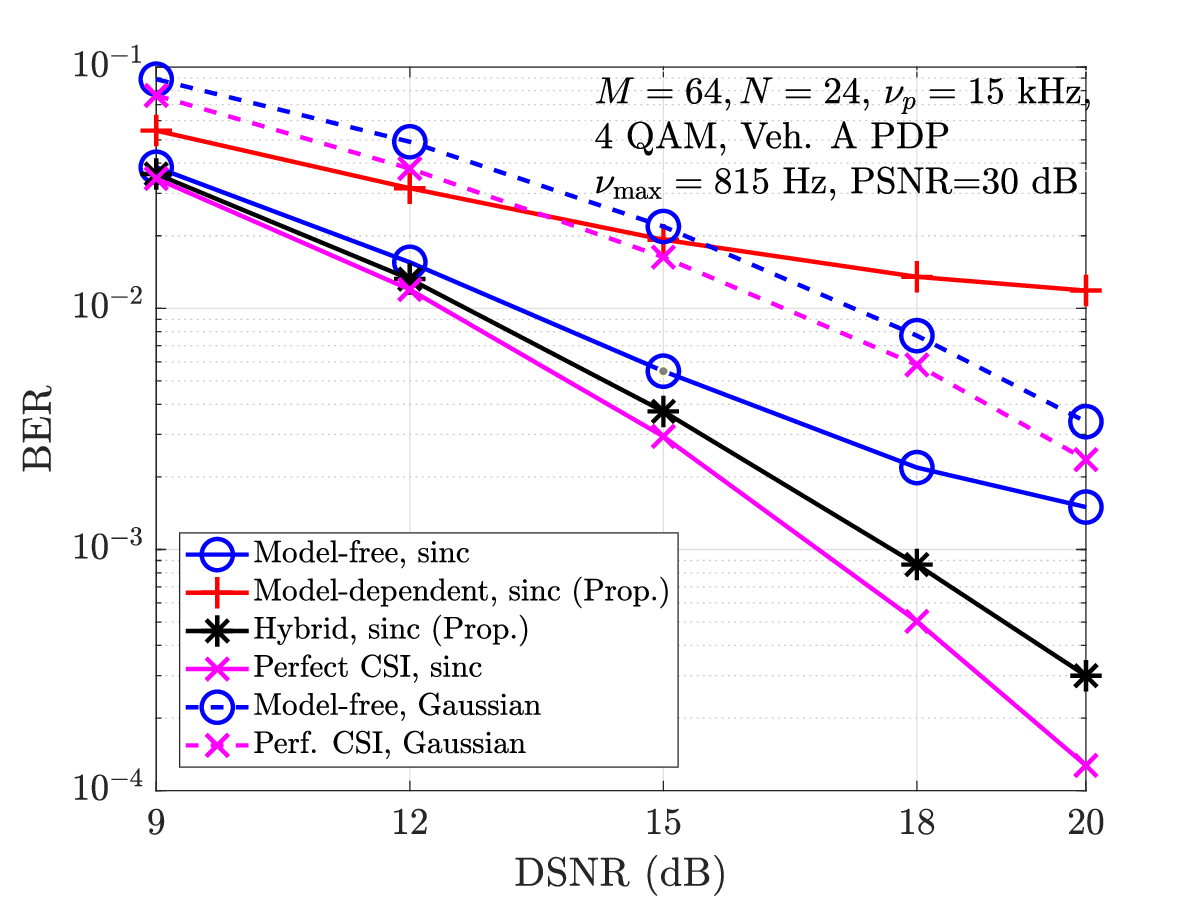}
     \caption{BER performance of the proposed hybrid estimation scheme as a function of DSNR for exclusive pilot frame at 30 dB PSNR.}
    \label{fig:ber_vs_dsnr_exclusive_pilot}
\end{figure}

\subsubsection{Detection performance}
Figure~\ref{fig:ber_vs_psnr_exclusive_pilot} shows the BER performance of the proposed hybrid estimation scheme as a function of PSNR for sinc and Gaussian filters at a fixed DSNR of 15 dB, and 
Fig.~\ref{fig:ber_vs_dsnr_exclusive_pilot} shows the BER performance as a function of DSNR at a fixed PSNR of 30 dB. For the sinc filter, at low PSNRs, the model-dependent approach achieves better BER performance, which is in agreement with the NMSE behavior observed earlier. As the PSNR increases, the BER of the model-free scheme improves, eventually surpassing that of the model-dependent scheme. The hybrid scheme follows a similar trend. Beyond a PSNR of about 15 dB, the hybrid scheme outperforms both the model-free and model-dependent schemes, demonstrating its advantage in the medium-to-high SNR range of interest. Also, for the Gaussian filter, the BER performance achieved with model-free estimate at 30 dB PSNR is close to that with perfect CSI, attributable to its excellent localization. On the other hand, although the sinc filter with perfect CSI provides much superior performance compared to that of the Gaussian filter with perfect CSI, sinc filter's performance with model-free estimate is far from its perfect CSI performance, attributable to its poor localization. Under these circumstances, the proposed hybrid estimation scheme brings the sinc filter's performance close to its perfect CSI performance. Also, it is seen that the use of the proposed model-dependent estimate alone is not competitive (due to its simplified nature), and hence
the proposed hybrid scheme is crucial to approach the perfect CSI performance of the sinc filter.

\begin{figure}
\centering
\includegraphics[width=0.875\linewidth]{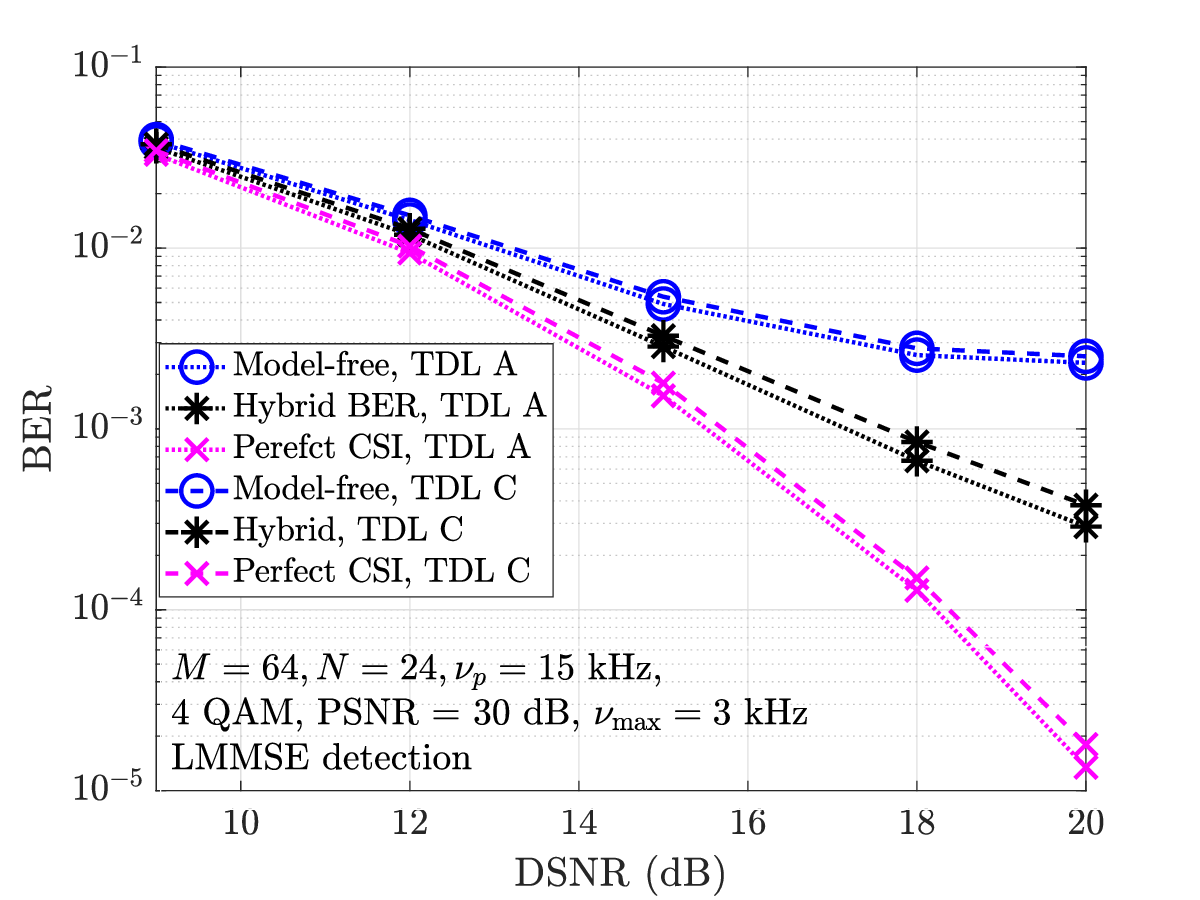}
\caption{BER performance of the proposed hybrid estimation scheme with sinc filter under 3GPP TDL-A and TDL-C channel models.}
\label{fig:TDL}
\end{figure}

Figure~\ref{fig:TDL} presents the BER performance of the proposed hybrid scheme with sinc filter under TDL-A and TDL-C channel models defined in 3GPP \cite{3gpp}, which have $P=23$ and $24$ paths, respectively. The urban macro delay profile with delay scaling factor $302$ ns and carrier frequency $f_c=15$ GHz (Table 7.7.3-2 in~\cite{3gpp}) is considered. For this $f_c$ and a maximum velocity of $220$ km/h, the maximum Doppler shift is about $3$ kHz. All other parameters remain the same as before. 
From Fig.\ref{fig:TDL}, it can be seen that, for both TDL-A and TDL-C channels, while the model-free scheme significantly falls short of the ideal (perfect CSI) performance, the proposed hybrid scheme is found to achieve substantially improved performance compared to the model-free scheme. For example, the proposed scheme achieves an SNR gain of about 4 dB at a BER of $2\times10^{-3}$ compared to the model-free scheme.

\begin{table*}[t]
\centering
\caption{Complexity of various estimation algorithms}
\resizebox{\linewidth}{!}{
\begin{tabular}{|c|c|c|}
\hline
\textbf{Estimation scheme}      & \textbf{No. of real additions} &\textbf{ No. of real multiplications}  \\ \hline
\textbf{Model-free  }           &   \(98M^2N^2+4MN\)                      & \(100M^2N^2+8MN\)                 \\ \hline  
\textbf{Model-dependent }(Prop.)  & 
\makecell{\(98M^2N^2+(6M+1)(6N+1)3\hat{P}_{\max}+2(\hat{P}_{\max}-1)\)\\
+\(4\hat{P}_{\max}^2MN+8\hat{P}_{\max}^3+4\hat{P}_{\max}MN-6\hat{P}_{\max}MN\)   }                     &
\makecell{\(100M^2N^2+(6M+1)(6N+1)9\hat{P}_{\max}+4\hat{P}_{\max}^2MN\)\\
\(+8\hat{P}_{\max}^3-8\hat{P}_{\max}^2+4\hat{P}_{\max}(MN-2)+(k_{\max}+1)(2l_{\max}+1)\)
}\\ \hline  
\textbf{Hybrid scheme }(Prop.) &  
\makecell{
\(98M^2N^2+\left((6M+1)(6N+1)-MN\right)3\hat{P}_{\max}+2(\hat{P}_{\max}-1)\)\\
\(4\hat{P}_{\max}^2MN+8\hat{P}_{\max}^3+4\hat{P}_{\max}MN-6\hat{P}_{\max}MN+4MN+(k_{\max}+1)(2l_{\max}+1)\)} &
\makecell{
\(100M^2N^2+\left((6M+1)(6N+1)-MN\right)9\hat{P}_{\max}\)\\
\(+4\hat{P}_{\max}^2MN-8\hat{P}_{\max}^3+4\hat{P}_{\max}(MN-2)+8MN\)
}
\\ \hline
\textbf{SBL} in \cite{off_grid}       &     
\makecell{
\(T_{\max}M^3N^3+98M^2N^2+T_{\max}2M_{\tau}N_{\nu}MN(3+4M_\tau N_\nu+4MN)\)\\
\(+(6M+1)(6N+1)3\hat{P}+2(\hat{P}-1)+M_\tau N_\nu\)
}
&     
\makecell{
\(T_{\max}M^3N^3+100M^2N^2+T_{\max}M_{\tau}N_{\nu}MN(1+M_{\tau}N_{\nu}+MN)\)\\
\(+(6M+1)(6N+1)9\hat{P}+2M_{\tau}N_{\nu}\)
}\\ \hline
\end{tabular}}
\label{tab:complexity}
\end{table*}

\subsubsection{Comparison with SBL-based model-dependent estimation}
\begin{figure}[t]
    \subfigure[BER vs DSNR.]{\label{fig:sbl_comp}\includegraphics[width=0.5\linewidth]{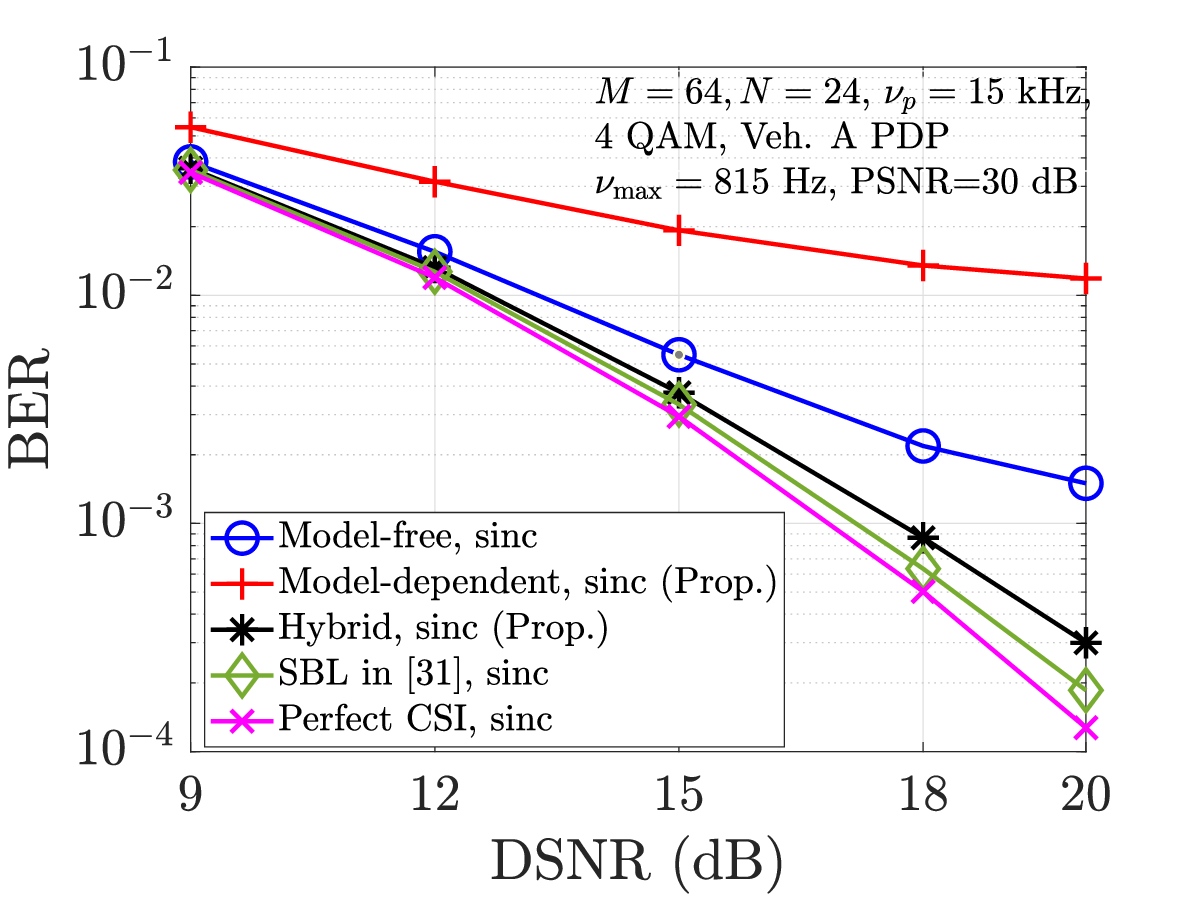}}
    \hspace{-2mm}
    \subfigure[Complexity.]{\label{fig:complexity}\includegraphics[width=0.5\linewidth]{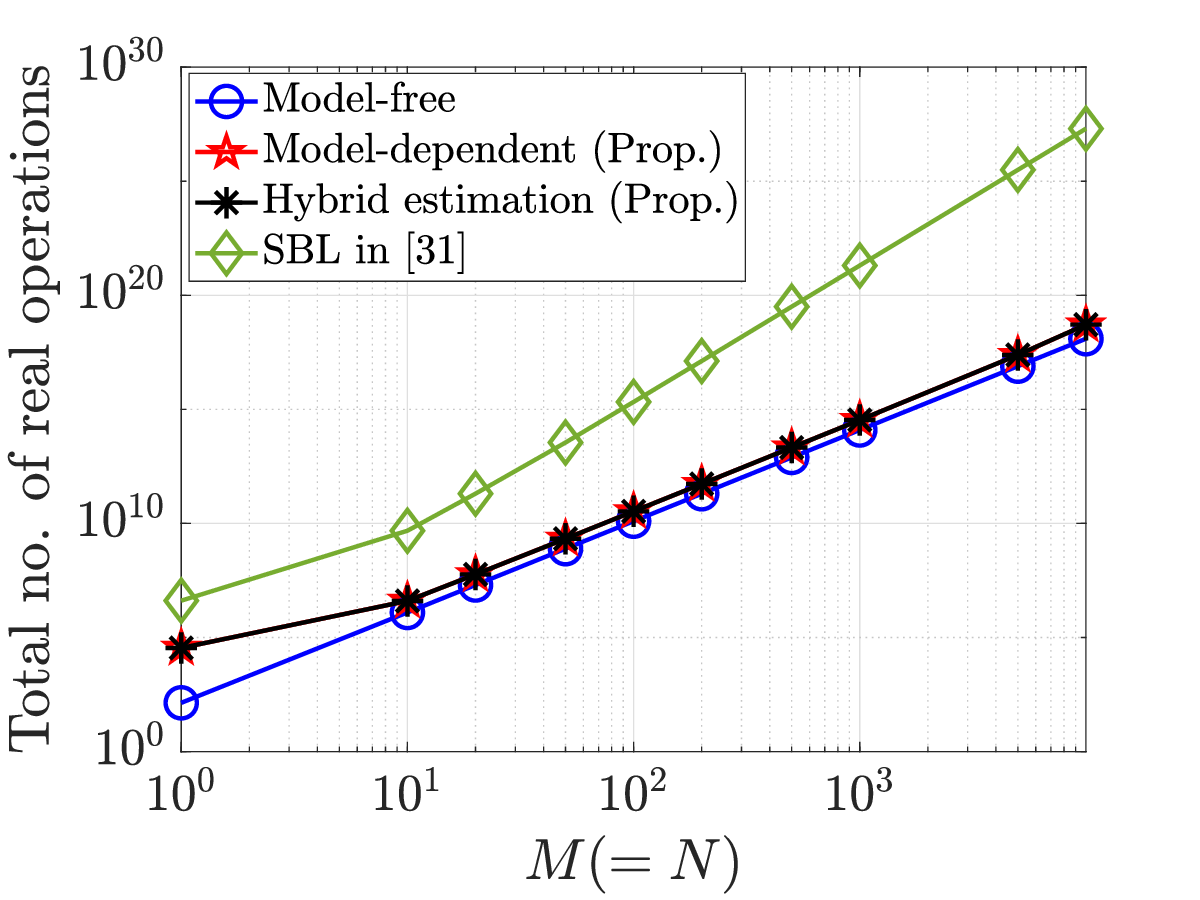}}
    \caption{BER performance and complexity (in number of real operations) of various estimation algorithms.}
    \label{fig:sbl}
\end{figure}

In Fig.~\ref{fig:sbl}, we present a BER and complexity comparison between the proposed hybrid estimation scheme and the SBL based model-dependent estimation in \cite{off_grid}. Figure~\ref{fig:sbl_comp} presents the BER comparison and Fig.~\ref{fig:complexity} presents the complexity comparison. Figure~\ref{fig:sbl_comp} shows the BER performance of the estimation schemes, namely, the proposed hybrid estimation scheme, the on-grid SBL algorithm in \cite{off_grid}, the proposed low-complexity estimation scheme, and the model-free estimation scheme. The performance with perfect CSI is also included as a benchmark. Other than the system parameters used for the proposed scheme, additional parameters specific to the on-grid SBL algorithm used are \cite{off_grid}: virtual grid parameters \(M_\tau = 11\), \(N_\nu = 14\), number of bins considered for estimation  \(M_T = M\), \(N_T = N\) (i.e., exclusive pilot frame), and maximum iteration count \(T_{\max} = 1000\). For reconstruction of the I/O relation matrix, the top \(\hat{P} = 15\) dominant estimated paths (in terms of energy)  are selected. All other parameters in the SBL algorithm are chosen to be the same as those in \cite{off_grid}. From Fig.~\ref{fig:sbl_comp}, it is observed that, among the considered estimation schemes, the SBL algorithm achieves the best BER performance (close to perfect CSI performance), closely followed by the proposed hybrid scheme. The difference in performance between the hybrid and SBL schemes is quite small.

{\em Complexity analysis:} The model-free scheme is the least expensive with a complexity  order of $\mathcal{O}(M^2N^2)$, which arises from the generation of the $M^2N^2$ entries of the $MN\times MN$ matrix $\hat{\boldH}$ (see \eqref{eqn:H_vec}) using the direct read-off $\hat{h}_{\eff}^{\free}$ values. The proposed hybrid scheme incurs this model-free complexity plus the additional computational complexity for estimating path gains and generating $\hat{h}_{\eff}^{\dep}$ in the region $\mathcal{F}_{\exc}^{\mathrm{c}}$. As summarized in Table~\ref{complexity}, this additional complexity scales linearly with $MN$ and cubically with the maximum number of estimated paths $\hat{P}_{\max}$. In contrast, the SBL algorithm incurs a significantly higher complexity of order $\mathcal{O}(T_{\max}M^3N^3)$, which arises from repeated matrix inversions of size $MN\times MN$ in each iteration, along with the computations required to generate $\hat{h}_{\eff}^{\dep}$ across the entire DD space. The overall complexity trends are evident from Table~\ref{tab:complexity} and Fig.~\ref{fig:complexity}. It is interesting to note that the proposed hybrid scheme is able to achieve performance comparable to SBL scheme at a significantly lower complexity.

\subsection{Performance with embedded pilot frame}
\subsubsection{Estimation accuracy}
\begin{figure}
    \centering
    \includegraphics[width=0.875\linewidth]{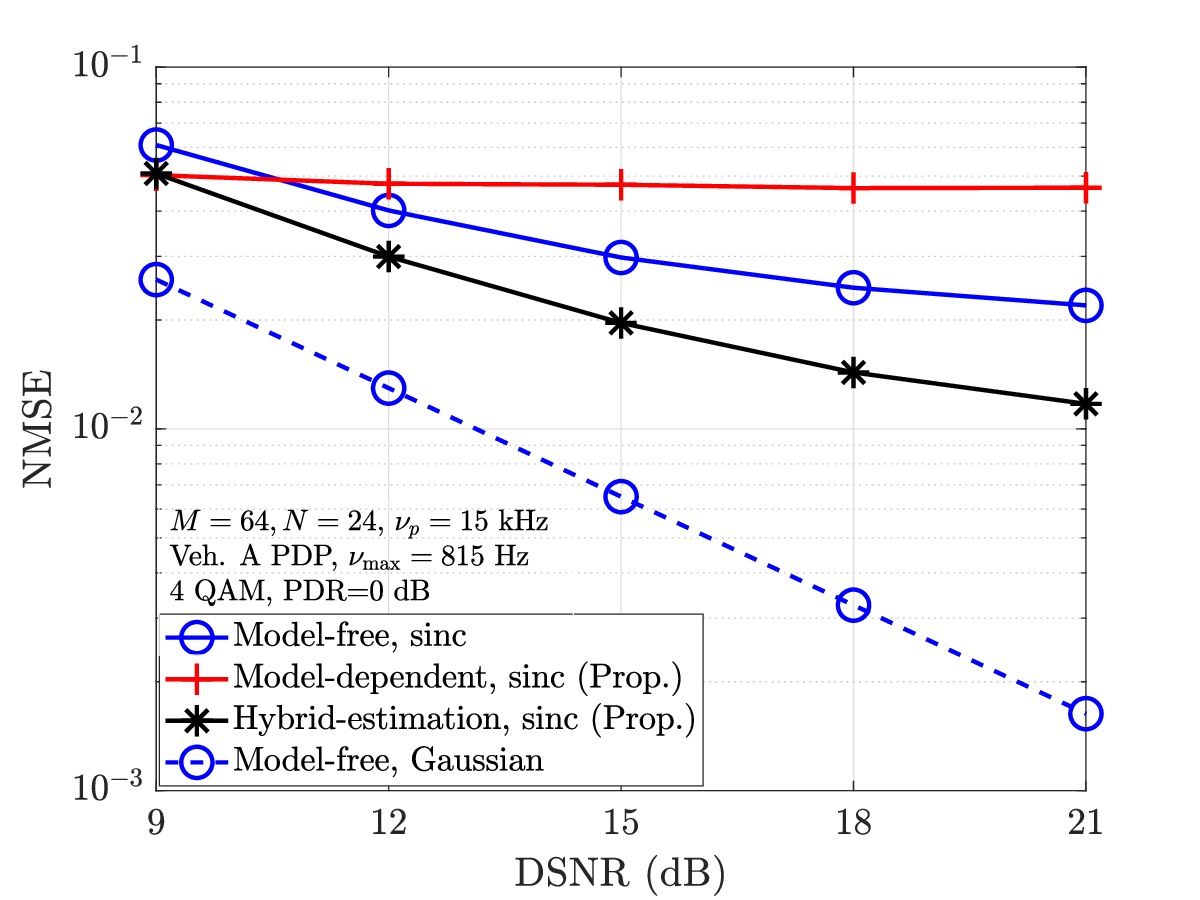}
    \caption{NMSE performance of the proposed hybrid estimation scheme as a function of DSNR for embedded pilot frame at 0 dB PDR.}
    \label{fig:nmse_embedded}
\end{figure}
Figure \ref{fig:nmse_embedded} presents the NMSE performance of the model-free, proposed model-dependent, and proposed hybrid estimation schemes as a function of DSNR at a PDR of 0 dB for sinc filter with embedded pilot frame. Model-free estimation performance with the Gaussian filter is also shown. The results demonstrate that while the performance of the low-complexity model-dependent approach floors as the DSNR increases, its integration within the proposed hybrid framework contributes to overall performance enhancement. Specifically, the hybrid scheme consistently outperforms both the model-free and proposed model-dependent approaches, highlighting the synergistic benefits of combining both approaches. The Gaussian filter is found to achieve very good estimation performance in the embedded pilot frame as well, because its better localization is able to effectively limit pilot-data interference.

\begin{figure}
    \centering
    \includegraphics[width=0.875\linewidth]{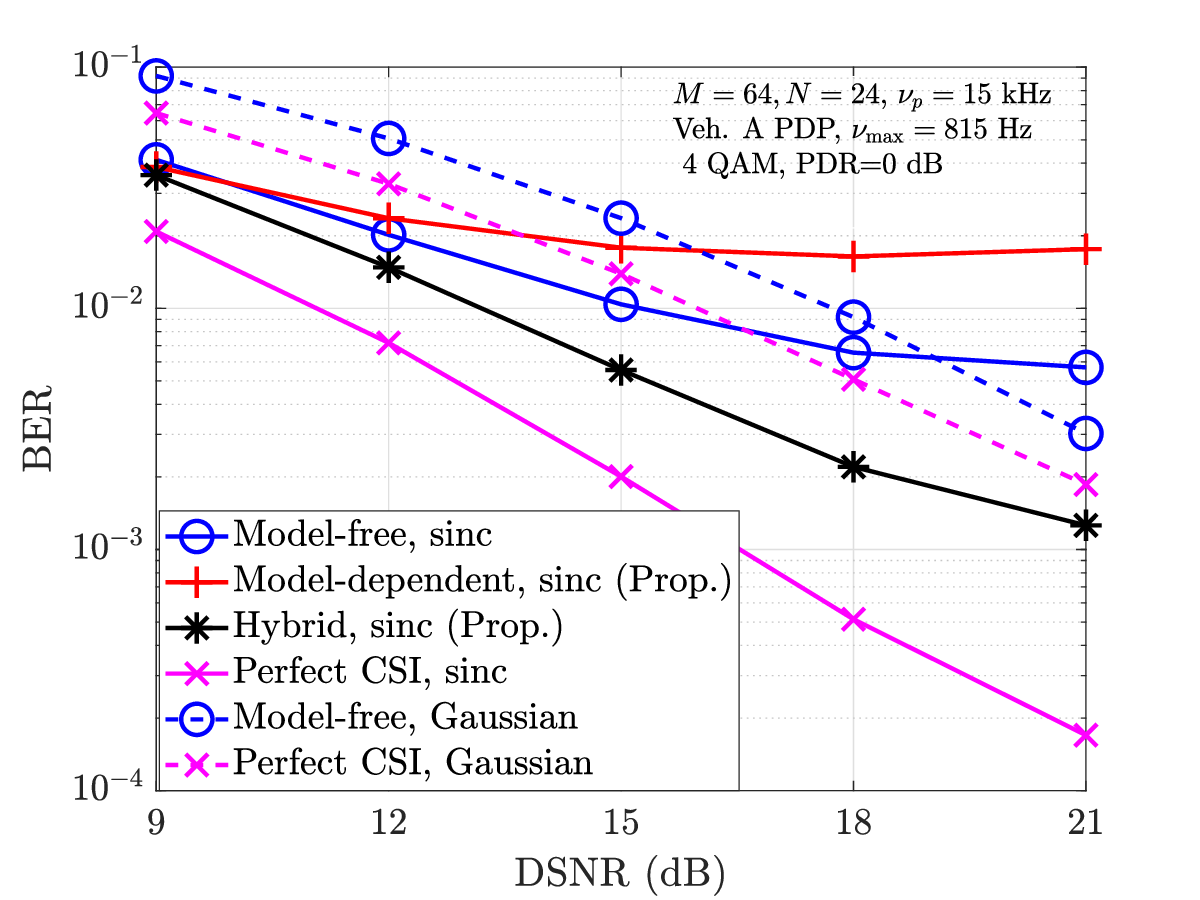}
    \caption{BER performance of the proposed hybrid estimation scheme as a function of DSNR for embedded pilot frame at 0 dB PDR.} 
    \label{fig:ber_embedded}
\end{figure}

\subsubsection{Detection performance}
Figure~\ref{fig:ber_embedded} shows the BER performance as a function of DSNR for sinc and Gaussian filters with embedded pilot frame at a fixed PDR of $0$ dB. Performance with perfect CSI is included as a reference benchmark. The model-free estimation yields good BER performance for Gaussian filter, performing close to its perfect CSI counterpart. However, with sinc filter, the model-free performance is poor, and, in some cases, even inferior to that of the Gaussian filter's model-free estimation performance, despite the fact that the perfect CSI performance of the sinc filter is superior compared to that of the Gaussian filter. This is attributed to the limited read-off region in the embedded frame, which is inadequate for the sinc filter. With the proposed hybrid estimation, however, the performance not only surpasses that of the model-free estimation but also exceeds the perfect CSI performance of the Gaussian filter. This highlights the effectiveness of the hybrid scheme in harnessing the Nyquist advantage of the sinc filter (nulls at DD sampling points), making it a good choice for low-complexity high-performance channel estimation. 
\begin{figure}
    \centering
    \includegraphics[width=0.875\linewidth]{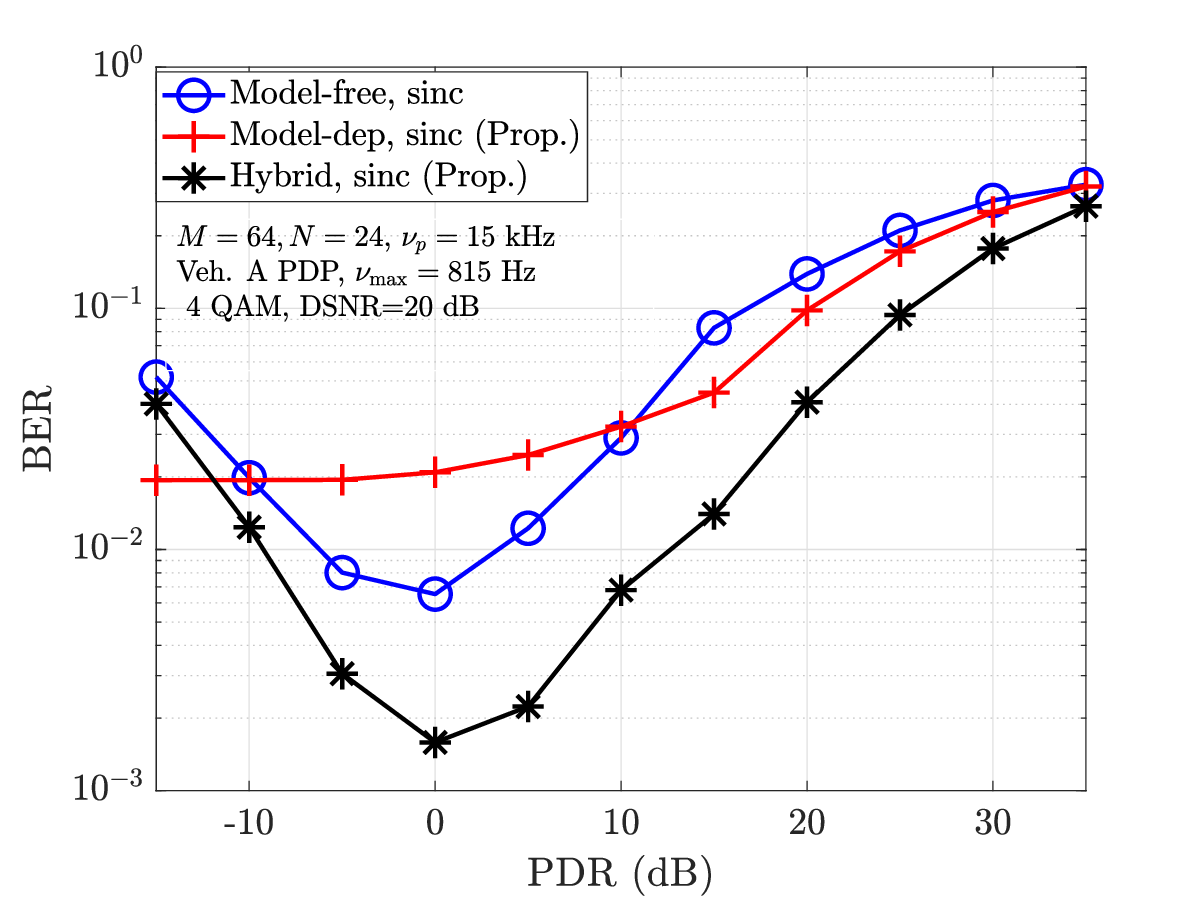}
    \caption{BER performance of the proposed hybrid estimation scheme as a function of PDR for embedded pilot frame at 20 dB DSNR.}
    \label{fig:ber_pdr}
\end{figure}

Figure~\ref{fig:ber_pdr} shows the BER performance of the model-free, proposed model-dependent, and proposed hybrid estimation schemes as a function of PDR for sinc filter with embedded pilot frame at a DSNR of 20 dB. A classical U-shaped curve is observed for the three schemes, i.e., both low and high PDRs give poor BER performance (because of poor estimation performance at low PDRs due to weak pilot power and poor detection performance at high PDRs due to increased pilot interference to data), and the performance is best at a certain optimum PDR in between. Among the three estimation schemes considered, the proposed hybrid scheme achieves the best performance. 

\section{Conclusions}
\label{conclusions}
In this work, we considered the problem of I/O relation estimation in Zak-OTFS systems. While the model-free approach offers simplicity and the ability to handle fractional delays and Dopplers, its performance is highly dependent on the localization characteristics of the pulse shaping filter. We observed that sinc pulses, though optimal under perfect CSI, perform poorly with model-free estimation due to their poor localization with high side lobes. Whereas Gaussian pulses, despite being sub-optimal in their main lobe characteristics, yield better model-free estimation performance due to their better localization with very low side lobes. To address this issue, we proposed a hybrid estimation framework that combines the simplicity of model-free read-off with a low-complexity model-dependent add-on. The hybrid approach was designed to extend the estimation beyond the limited region captured by model-free estimation, thereby compensating for its shortcomings. We developed this framework for both exclusive and embedded pilot configurations and analyzed the NMSE and BER performance.
Our simulation results showed that the proposed hybrid scheme achieves lower NMSE and improved BER compared to model-free estimation alone, especially with sinc pulses. The proposed scheme was shown to achieve performance comparable to that of the SBL algorithm at a significantly lower complexity. These results show that the hybrid estimation framework can bring the BER performance towards the perfect CSI performance of the sinc filter, enabling Zak-OTFS to fully exploit the benefits of optimal DD pulse shaping in high-mobility channels. 
{\color{black} In this work, we have considered exclusive and embedded pilot frames which incur throughput loss due to pilot overhead in a frame. Zak-OTFS with superimposed pilot frames (where a point pilot spread in the DD domain is superimposed on data in order to avoid throughput loss due to pilot overhead) can be considered as one possible direction for future work. Investigation of learning-based techniques for achieving robust I/O relation estimation can be another promising line of future research. 
}

\appendices
\section{Proof of Eqn. \eqref{eqn:sys_model_non_vec}}
Ignoring the noise and by substituting \eqref{eqn:x_DD} in \eqref{eqn:y_DD},
\begin{align}
    &y_{\DD}^{w_{\rx}}(\tau,\nu)=h_{\eff}(\tau,\nu)\twistconvol x_{\DD}(\tau,\nu)=\nonumber\\
    &\iint h_{\eff}(\tau',\nu')x_{\DD}(\tau-\tau',\nu-\nu')
    e^{j2\pi\nu'(\tau-\tau')}d\tau'd\nu'\nonumber\\ =&\iint\sum_{m,n\in\mathbb{Z}}\sum_{k=0}^{M-1}\sum_{l=0}^{N-1}x[k,l]e^{j2\pi\frac{nl}{N}}e^{j2\pi\nu'(\tau-\tau')}\nonumber\\
    &\hspace{2mm}\delta\left(\nu-\nu'-\tfrac{(l+mN)\nup}{N}\right)\delta\left(\tau-\tau'-\tfrac{(k+nM)\taup}{M}\right)d\tau'd\nu'\nonumber\\
    =&\sum_{m,n\in\mathbb{Z}}\sum_{k=0}^{M-1}\sum_{l=0}^{N-1}x[k,l]e^{j2\pi\frac{nl}{N}}\hspace{-1mm}\iint \hspace{-1mm}h_{\eff}(\tau',\nu')e^{j2\pi\nu'(\tau-\tau')}\nonumber\\
    &\hspace{3mm}\delta\left(\tau-\tau'-\tfrac{(k+nM)\taup}{M}\right)\delta\left(\nu-\nu'-\tfrac{(l+mN)\nup}{N}\right)d\tau'd\nu'\nonumber\\
    =&\sum_{m,n\in\mathbb{Z}}\sum_{k=0}^{M-1}\sum_{l=0}^{N-1}x[k,l]e^{j2\pi\frac{nl}{N}}e^{j2\pi\left(\frac{(k+nM)\taup}{M}\right)\left(\nu-\frac{(l+mN)\nup}{N}\right)}\nonumber\\
    &\hspace{3mm}h_{\eff}\left(\tau-\tfrac{(k+nM)\taup}{M},\nu-\tfrac{(l+mN)\nup}{N}\right)\nonumber.
\end{align}
Upon sampling the signal at $\left(\frac{k'\taup}{M},\frac{l'\nup}{N}\right)$, the received DD signal becomes
\begin{align*}
    &y_{\DD}^{w_{\rx}}[k',l']=\nonumber\\  &\sum_{m,n\in\mathbb{Z}}\hspace{-1mm}\sum_{k=0}^{M-1}\hspace{-1mm}\sum_{l=0}^{N-1}\hspace{-1mm}x[k,l]e^{j2\pi\frac{nl}{N}}\hspace{-1mm}e^{j2\pi\Big(\big(\frac{(k+nM)\taup}{M}\big)\left(\frac{l'\taup}{M}\frac{(l+mN)\nup}{N}\right)\Big)}\nonumber\\
    &\hspace{3mm}h_{\eff}\left(\tfrac{k'\taup}{M}-\tfrac{\left(k+nM\right)\taup}{M},\tfrac{l'\nup}{N}-\tfrac{(l+mN)\nup}{N}\right)\nonumber\\
=&\sum_{m,n\in\mathbb{Z}}\sum_{k=0}^{M-1}\sum_{l=0}^{N-1}x[k,l]e^{j2\pi\frac{nl}{N}}e^{j2\pi\frac{(l'-l-mN)(k+nM)}{MN}}\nonumber\\
    &\hspace{3mm}h_{\eff}[{k'-k-nM,l'-l-mN}]. 
\end{align*}
$\hfill\blacksquare$

\end{document}